\documentclass[12pt]{iopart}

\usepackage{epsfig}
\usepackage{ifpdf}
\usepackage{graphicx}

\usepackage{amssymb,lineno,mathbbol,upgreek}

\usepackage[tight]{subfigure}

\usepackage{sidecap}

\usepackage{bbm}
\usepackage{slashed}
\usepackage{calligra}
\DeclareMathAlphabet{\mathcalligra}{T1}{calligra}{m}{n}
\DeclareFontShape{T1}{calligra}{m}{n}{<->s*[2.2]callig15}{}

\usepackage{calligra}
\DeclareMathAlphabet{\mathcalligra}{T1}{calligra}{m}{n}
\DeclareFontShape{T1}{calligra}{m}{n}{<->s*[2.2]callig15}{}
\usepackage{enumitem}
\begin{document}


\title[Superfluid fluctuations and emergent theories]{The nonlinear Dirac equation in Bose-Einstein condensates: Superfluid fluctuations and emergent theories from relativistic linear stability equations}

\author{L H Haddad$^1$ and Lincoln D Carr$^{1,2}$}
\address{$^1$Department of Physics, Colorado School of Mines, Golden, CO 80401,USA   \\$^2$Physikalisches Institut, Universit\"at Heidelberg, D-69120 Heidelberg, Germany}

\ead{\mailto{laith.haddad@gmail.com}, \mailto{lcarr@mines.edu}}

\begin{abstract}
We present the theoretical and mathematical foundations of stability analysis for a Bose-Einstein condensate (BEC) at Dirac points of a honeycomb optical lattice. The combination of s-wave scattering for bosons and lattice interaction places constraints on the mean-field description, and hence on vortex configurations in the Bloch-envelope function near the Dirac point. A full derivation of the relativistic linear stability equations (RLSE) is presented by two independent methods to ensure veracity of our results. Solutions of the RLSE are used to compute fluctuations and lifetimes of vortex solutions of the nonlinear Dirac equation, which include Anderson-Toulouse skyrmions with lifetime $\approx 4$ seconds. Beyond vortex stabilities the RLSE provide insight into the character of collective superfluid excitations, which we find to encode several established theories of physics. In particular, the RLSE reduce to the Andreev equations, in the nonrelativistic and semiclassical limits, the Majorana equation, inside vortex cores, and the Dirac-Bogoliubov-de Gennes equations, when nearest-neighbor interactions are included. Furthermore, by tuning a mass gap, relative strengths of various spinor couplings, for the small and large quasiparticle momentum regimes, we obtain weak-strong Bardeen-Cooper-Schrieffer superconductivity, as well as fundamental wave equations such as Schr\"odinger, Dirac, Klein-Gordon, and Bogoliubov-de Gennes equations. Our results apply equally to a strongly spin-orbit coupled BEC in which the Laplacian contribution can be neglected. 
\end{abstract}


\submitto{\NJP}

\maketitle

\section{Introduction}

Two contemporary themes in the study of cold atomic gases are the creation of new exotic forms of quantum matter, and quantum simulations of systems already present in nature~\cite{Unruh2007,bloch2008,Lewenstein2012}. By tuning the parameters for a collection of atoms and lasers one may address problems in quantum many-body systems or in high-energy physics~\cite{Volovik2003}. In the first case degeneracy, quantum correlation, and entanglement are essential ingredients, whereas the latter case usually focuses on low-energy fluctuations of systems where a macroscopic fraction of particles reside in a single quantum state, often amenable to Landau descriptions. The versatility of Bose-Einstein condensates (BECs) allows the freedom to specify the geometry and topology of the order parameter to suit a particular purpose. For example, spinor BECs provide one way to realize order parameters with large symmetry groups, and hence exotic topologies~\cite{Ketterle2003,Cornell1999,Holland1999,Cornell:2004,kasamatsu2005,Ueda2008,Kawaguchi2011,Kurn2013,Fetter2014,Frantz2015,Ohberg2014}. It follows that the inclusion of spin-orbit coupling in such systems increases their complexity and introduces topological order~\cite{Zhai2012,Wu2013,Galitski2013,Ueda2014,Borgh2014,Zhao2015}, a distinct classification for the order parameter. However, in order to access interesting physics and to simulate new regimes it may be necessary to extend beyond the usual notion of stability to metastable non-ground-state or non-equilibrium BECs.

In this article, we develop some of the fundamentals underlying non-ground-state BECs in quasi-two-dimensional (quasi-2D) honeycomb lattices and the associated long-wavelength emergent theories~\cite{Carr2000,Sengstock10,Sengstock11,Tarruell2012,Salasnich2004,haddad2009,Ablowitz2012,Ablowitz2013}. We focus in particular on superfluid fluctuations in the presence of Dirac points from a semiclassical perspective and by including lowest-order quantum effects. Quantum fluctuations are determined by solving the partial differential equations which describe dynamics of the low-energy modes for an arbitrary condensate profile. These equations are Lorentz invariant and comprise a relativistic generalization of the Bogoliubov-de Gennes equations (BdGE); thus we call them \emph{relativistic linear stability equations} (RLSE). The RLSE provide a means of calculating vortex stabilities, yet their versatility extends beyond stability calculations to simulating a large number of established theories in addition to some exotic ones. This is because quasiparticles in BECs with inherent relativistic structure (e.g., linear dispersion, $\mathcal{CPT}$ invariance, multicomponent order parameter, etc.) can be tuned to have linear or quadratic dispersion with a zero or finite gap coupled to a condensate reservoir with a large number of possible internal symmetries.$^{\footnotemark[1]}$ \footnotetext[1]{Relativistic effects occur in photonic systems as well. See for example Refs.~\cite{Dreisow2010,Szameit2011,Purvis2013,Rubino2012}. Interesting dynamics in the merging of Dirac points for fermionic systems have also been investigated~\cite{1Lim2012,2Lim2012}}In short, ease of construction and manipulation of BECs with quasi-relativistic dispersion and characteristically low sound speeds (on the order of centimeters per second) present ideal environments for simulating high energy phenomena.

Moreover, the ``no-node'' theorem originally proposed by Feynman~\cite{Feynman1972}, which constrains conventional  BECs, is circumvented for non-ground state (metastable) systems and in the case of spin-orbit coupling, as the order parameter in these systems is generally not positive-definite~\cite{CWu2009,Zhou2013}. This property is a fundamental feature of quasi-relativistic condensed matter systems. In particular, lifting the ``no-node'' theorem restriction leads to time-reversal symmetry breaking, which allows for exotic bosonic systems such as p-wave superfluids~\cite{ohmi2010}, chiral Bose liquids~\cite{Liu2014}, complex unconventional BECs in high orbital bands of optical lattices~\cite{Wu2011}, and BECs with repulsive interactions that support bright solitons and vortices as well as skyrmions~\cite{haddadcarrsoliton1,Haddad2015,Wu2013}. We point out that our system is identical to a quasi-2D BEC with spin-orbit coupling in either the long-wavelength limit or the strong tunable spin-orbit coupled limit, provided the interactions are also chosen to retain only the intra-component terms. To map to the strong spin-orbit coupled limit, however, the strength of the spin-orbit coupling term must be much larger than the quadratic term but still below the quantum critical point separating the spin-balanced and spin-polarized ground states~\cite{Zhang2013}.

Our results focus on three main topics: \emph{1) physical parameters and constraints}; \emph{2) linear stability of vortices}; and \emph{3) emergent theories}. First, the physical parameters and necessary constraints to construct a non-ground-state condensate at Dirac points are explained in detail. The BEC is tightly confined in one direction and loosely confined in the other two directions. More precisely stated, magnetic trapping along the $z$-direction is such that excitations along this direction have much higher energy, by at least an order of magnitude, compared to the lowest excitations in the $x$ and $y$-directions. Thus, an important step is to calculate the precise renormalization of all relevant physical parameters when transitioning from the standard 3D BEC to a quasi-2D system. In addition to this step we also account for renormalization due to the presence of the optical lattice potential which introduces an additional length scale from the lattice constant. We point out that microscopically the BEC obeys the three-dimensional nonlinear Schr\"odinger equation and we consider temperatures well below the BKT transition energy associated with two-dimensional systems. Nevertheless, throughout our work we often use ``2D'' for brevity, keeping in mind the quasi-2D picture. Condensation at Dirac points of the honeycomb lattice requires additional techniques beyond ordinary condensation, which we have detailed in our previous work~\cite{Haddad2012}. In addition to the fields needed to construct the lattice one requires a resonant field which provides the time-dependent potential to ``walk'' atoms from the ground state (zero crystal momentum) to the Dirac point. The result is a transient configuration since a macroscopically occupied nonzero Bloch mode is not in thermodynamic equilibrium.

Care must be taken when transferring atoms from the ground state to a Dirac point in order to minimize depletion out of the condensate. In general, one might expect some dissipation to occur due to secondary interactions within the condensate, and between condensed atoms and the lattice, quantum fluctuations, and thermal excitations, the latter two comprising the normal fluid. However, at the mean-field level repulsive atomic interactions within the condensate itself produce a single Hartree term which just shifts the total energy upward without causing additional depletion. Lattice effects are accounted for completely through the band dispersion, since we are not considering the presence of disorder or artificial impurities. Moreover, we consider only the zero-temperature case. There is certainly finite leakage into energetic modes lower as well as higher than the condensate energy. However, such losses can only occur in the presence of higher-order dissipative terms in the Hamiltonian. In this article we restrict our analysis to the effects of first-order quantum corrections and apply our results to the special case of vortex background.

The second major topic in this article addresses linear stability of vortices near a Dirac point. We first provide a detailed derivation of the RLSE then solve them for vortex solutions of the nonlinear Dirac equation~\cite{Haddad2015}. The resulting eigenvalues determine the characteristic lifetimes of each vortex type. Solutions of the RLSE are inherently massless Dirac spinors with components that couple only through the Dirac kinetic terms. For a vortex background, RLSE solutions describe the quantum density and phase fluctuations near the vortex core. Physically, these are local undulations in the density profile, rigid translations of the vortex itself, and fluctuations in the speed of rotation. Although the latter is topologically protected, at the mean-field level quantum effects introduce small admixtures of different winding numbers into the vortex. These admixtures, which take the form of phase fluctuations, comprise the Nambu-Goldstone modes of the system. Near the vortex core they appear as bound states, the lowest of which are zero-energy modes (zero modes): static modes with zero energy associated with spatial translations of the center of the vortex. From a symmetry perspective, zero modes account for the fact that a vortex breaks the translational and rotational symmetry of an otherwise uniform system. We will address the various modes in generality when we discuss the associated reductions of the RLSE to other well known equations.

\begin{figure}[h]
\centering
\hspace{0pc} \includegraphics[width=.95\textwidth]{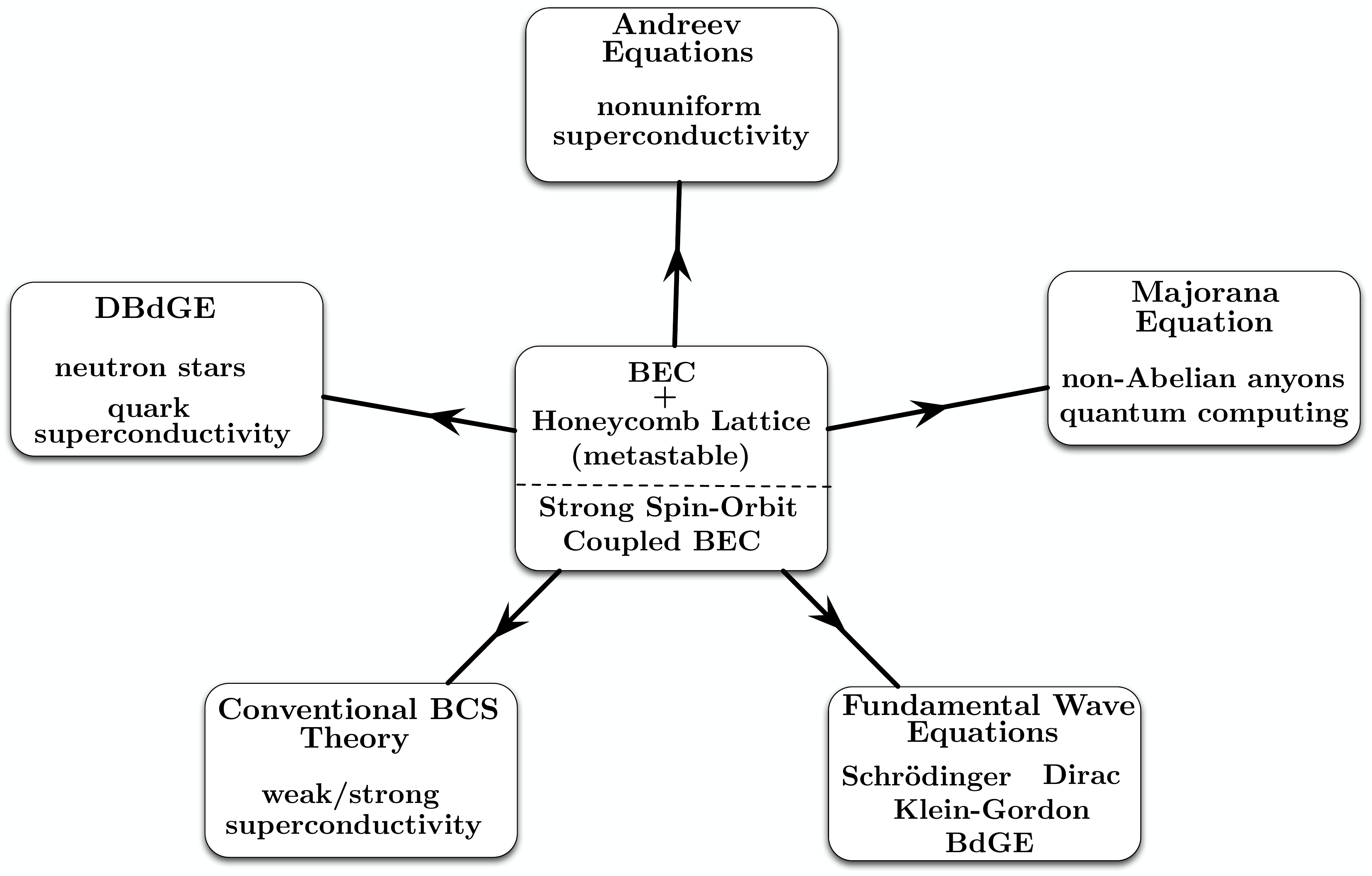} 
\hspace{0pc} \caption{(color online)     \emph{Schematic overview for applications of the RLSE}. The RLSE are derived for either a metastable BEC at Dirac points of a honeycomb optical or for a strong spin-orbit coupled BEC. Various reductions of the RLSE or augmentation by the addition of a mass gap and nearest-neighbor interactions (lattice case), or general spinor couplings (spin-orbit coupled case), leads to five categories of equations from different areas of physics. } 
\label{RLSEOverview}
\end{figure}

Our work culminates with the connection to several other important areas of physics including relativistic Bardeen-Cooper-Schrieffer (BCS) theory. In addition to continuous space-time dependence, quasiparticle solutions of the RLSE are labeled by two indices associated with the lattice pseudospin valley and the particle-hole structure analogous to Nambu space from BCS theory. In order to avoid confusion we will refer to these as valley and Nambu indices, and reserve the particle-hole terminology to distinguish between the two states in either the valley or Nambu space. The RLSE are formulated to describe excitations in a  repulsive Bose gas but can be reinterpreted as excitations in a theory comprised of attractive particles upon pseudospin valley particle-hole exchange. This symmetry is a consequence of the combined symmetry of charge conjugation ($\mathcal{C}$), parity transformation ($\mathcal{P}$), and time reversal ($\mathcal{T}$), which is fundamentally related to the structure of the Dirac operator. Retaining a mass gap, an intermediate step in Dirac-point condensation~\cite{Haddad2012}, and adding atomic interactions between nearest-neighbor lattice sites~\cite{Rossini2012} extends the RLSE to the Dirac-Bogoliubov-de Gennes equations (DBdGE) \cite{Capelle1999-1,Capelle1999-2,Ohsaku1,Ohsaku2}, provided valley particles and holes are interchanged. This connection is significant as DBdGE are required for a broader description of superconductivity beyond the standard BCS formalism, particularly for superconductors with a high Fermi velocity. Indeed, a relativistic formulation of BCS becomes important for elements with large atomic number ($Z \ge 40$), in neutron stars where superfluidity is expected to play a major role in ``glitches''~\cite{Hewish1994,Sauls1989,Anderson2007}, and in color superconductivity where the strong nuclear force provides the attraction between fermions~\cite{Bailin1984}. In the nonrelativistic and semiclassical limits the RLSE reduce to the Andreev equations. These equations were originally formulated to address physics of nonuniform superconductors, for instance a type-I superconductor near a normal-superfluid interface or a vortex in a type-II superconductor~\cite{Andreev1964,Ketterson1999}. Interestingly, we find that the RLSE reduce to the Majorana equation inside the core of NLDE vortices. From a fundamental standpoint the Majorana equation describes relativistic fermions that are their own anti-particle~\cite{Majorana1937}. In condensed matter systems, and in particular our problem, finite-energy phase fluctuations inside the core of an NLDE vortex connect smoothly to Majorana zero modes. This is significant as Majorana zero modes are presently of great interest in such fields as topological quantum computation~\cite{Nayak2008}, topological insulators~\cite{Kane2008}, and more generally in the study of non-Abelian anyons and fractional statistics~\cite{Leinaas1977,Wilczek1982,Volovik1999,Stern2008}. Figure~\ref{RLSEOverview} provides a schematic overview of some of the theories and physical regimes encapsulated in the RLSE.

This article is organized as follows. In Sec.~\ref{Constraints}, we discuss physical parameters, constraints, and regimes. In Sec.~\ref{Excitations}, we analyze superfluid excitations for a Bose gas in the honeycomb lattice from a semiclassical perspective. Section~\ref{RLSE} contains two derivations of the RLSE according to paths dictated by two possible orderings of the tight-binding and continuum limits of lattice Bloch functions. In Sec.~\ref{Stability}, stability analysis is performed for vortex solutions of the nonlinear Dirac equation by solving the RLSE for quasiparticle functions and eigenvalues. In Sec.~\ref{Reductions}, we examine several reductions of the RLSE to other well known equations. We map the RLSE to the equations for relativistic BCS theory and demonstrate the non-relativistic limit to standard BCS theory. In Sec.~\ref{Conclusion}, we conclude.

\section{Renormalized Parameters and Physical Constraints}
\label{Constraints}

To obtain the correct renormalized parameters for the NLDE we proceed by two steps. First, we follow the transformation of the 3D NLSE parameters as we reduce to the 2D NLSE. Second, we take the long-wavelength limit of the 2D theory at the Dirac point to get the NLDE, which induces a second renormalization of the parameters. 

\subsection{Transition from 3D to 2D nonlinear Schr\"odinger equation}

A BEC comprised of $N$ atoms of mass $M$ is described by a wavefunction $\psi({\bf r},t)$ which solves the time-dependent nonlinear Schr\"odinger equation. The single-particle density is defined as $|\psi({\bf r}, t)|^2$, the BEC density $\rho({\bf r},t)^2 \equiv N |\psi({\bf r}, t)|^2$, and the phase is $\phi \equiv \mathrm{arg}[\psi({\bf r}, t)]$, with the superfluid velocity given by ${\bf v}_s \equiv \hbar {\bf \nabla}\phi/M$. The two-particle interaction strength is $g= 4 \pi \hbar^2 a_s/M$ and the healing length is $\xi = 1/\sqrt{8 \pi \bar{n} a_s}$, where $a_s$ is the $s$-wave scattering length for binary collisions between atoms. We take $a_s>0$ so that $g>0$, i.e., we consider only repulsive interactions, leaving attractive interactions for future studies. Throughout our work, we treat the case of an axisymmetric system associated with a harmonic trapping potential with two large dimensions described by  a radius $R= \sqrt{x^2+y^2}$, and a small dimension transverse to the plane described by the length $L_z$. The average density which appears in $\xi$ is then $\bar{n} \equiv N/(\pi R^2 L_z)$. Note that $\psi({\bf r},t)$ has dimensions of length$^{-3/2}$ so that $g$ has dimensions of energy$\times$length$^{3}$. Another important quantity is the speed of sound in the condensate, which is defined as $c_s = \sqrt{g \bar{n}/M}$.

Transforming to the 2D regime requires that $a_s \ll L_z \lesssim \xi$~\cite{Petrov2000,Hadzibabic2011}, which ensures that the condensate remains in the ground state in the transverse direction, and $L_z \ll R$, which ensures that excitations along the plane have much lower energy than those in the transverse direction. The wavefunction can then be separated into longitudinal and transverse modes, following similar arguments as in Ref.~\cite{Carr2000}
\begin{eqnarray}
\psi({\bf r}, t) = ( A L_z)^{-1/2} f(x,y) h(z) e^{-i \mu t/\hbar} \, ,
\end{eqnarray}
where $f(x,y)$ and $h(z)$ are the dimensionless spatial functions that describe the longitudinal and transverse normal modes, respectively, and $\mu$ is the chemical potential. Projecting onto the ground state of the transverse dimension $h_\mathrm{gs}(z)$, gives us an effectively 2D wave equation. In the case where $L_z \sim \xi$, $h_\mathrm{gs}(z)$ is just the ground state of the one-dimensional particle-in-a-box solution~\cite{Carr2000}, we then have $h_\mathrm{gs}(z) =  \sqrt{2}\,  \mathrm{sin}(\pi z/ L_z)$. It may be convenient to express $L_z$ and $R$ in terms of the trap frequencies $\omega_x, \omega_y$, and $\omega_z$, in which case we may write $L_{z} = \left(\hbar /M \omega_{z}\right)^{1/2}$, $R=\sqrt{ \hbar  M^{-1} (1/\omega_{x} + 1/ \omega_{y})}$. The transformation is then completed by defining the renormalized 2D chemical potential and interaction as
\begin{eqnarray}
\mu_\mathrm{2D} \equiv   \mu + \frac{\hbar^2 \pi^2  }{2 M L_z^2 } \; , \;\;\; \; g_\mathrm{2D} \equiv  \frac{3}{2}   \frac{g}{L_z} \, . \label{g2D} 
\end{eqnarray}
Through this process, we obtain a reduction of the 3D nonlinear Schr\"odinger equation to a 2D form with renormalized chemical potential and interaction given by Eq.~(\ref{g2D}). The 2D renormalized average density can be related to the 3D average density using the transverse oscillator length or frequency
\begin{eqnarray}
\bar{n}_\mathrm{2D} \equiv \frac{N}{A} =  L_z \, \bar{n}  =  \left( \frac{\hbar}{M \omega_{z}} \right)^{1/2} \! \! \!\bar{n} \, . \label{nbar2D}
\end{eqnarray}
Using this definition and the 2D single-particle wavefunction, $\psi(x,y) = A^{-1/2} f(x,y)$, we can write the 2D condensate density as $\rho_\mathrm{2D}(x,y) = N \,  |\psi(x,y)|^2$. The 2D renormalized healing length can also be constructed which we find acquires only an extra numerical factor
\begin{eqnarray}
\xi_\mathrm{2D}\,  \equiv\,    \left( \frac{2}{3} \right)^{1/2}\!\!\!\!\!\! \frac{1}{\sqrt{8 \pi \bar{n} a_{s} }}  \, = \,   \left( \frac{2}{3} \right)^{1/2}\!\!\!\!\!\!  \xi \, .
\end{eqnarray}
Similarly, we find the 2D speed of sound to be $c_{s \mathrm{2D}} = \sqrt{g_\mathrm{2D} \bar{n}_\mathrm{2D}/M } =    \left( 3/2 \right)^{1/2} \!\! c_s$. It is important to keep track of the effect of the reduced dimensionality on the dimensions of the constants: $\psi(x,y)$ now has dimensions of length$^{-1}$, $g_\mathrm{2D}$ has dimensions energy$\times$length$^2$, and $\bar{n}_\mathrm{2D}$ has dimensions length$^{-2}$.

\subsection{Derivation of nonlinear Dirac equation from 2D nonlinear Schr\"odinger equation.}

The derivation of the nonlinear Dirac equation begins with the second quantized Hamiltonian for a 2D system with the bosonic field operators $\hat{\psi} \equiv \hat{\psi}({ \bf r},t) = \hat{\psi}(x,y,t)$ obeying bosonic commutation relations in the Heisenberg picture. We then expand in terms of Bloch states belonging to A or B sites of the honeycomb lattice which breaks up the bosonic field operator into a sum over the two sublattices. The spatial dependence in this expansion is encapsulated in the exponential Bloch wave and the Wannier functions $w(x,y)$ which are then integrated out leaving only number-operator terms in the form of a \emph{Dirac-Hubbard Hamiltonian} (a detailed derivation is presented in Ref.~\cite{haddad2009}) 
\begin{eqnarray}
\hspace{-5pc}  \hat{H} =   -t_h\sum_{<A,B>} \left[\hat{a}^{\dagger} \hat{b}\,e^{i { \bf k}\cdot ({ \bf r}_A- {\bf r}_B) } +  \hat{b}^{\dagger} \hat{a}\,e^{-i {\bf k}\cdot ( {\bf r}_A- {\bf r}_B) }\right] + \frac{U}{2}  \sum_A    \hat{a}^{\dagger} \hat{a}^{\dagger}  \hat{a}\hat{a} + \frac{U}{2}  \sum_B    \hat{b}^{\dagger} \hat{b}^{\dagger}  \hat{b}\hat{b} \, .  \label{DiracHubbard}
 \end{eqnarray}
The bracketed A and B summation index signifies a sum over nearest-neighbor A and B sites, accounting for inter-sublattice hopping through the individual sublattice creation and destruction operators $\hat{a}^{\dagger}$,$\hat{b}^{\dagger}$ and $\hat{a}$,$\hat{b}$, respectively, with an accumulated phase which depends on the separation vectors ${\bf r}_{AB} \equiv { \bf r}_A- {\bf r}_B$. The interaction terms appear quartic in operators of the same site, as contact interactions are local and do not couple the A and B sublattice. Note that $t_h$ and $U$ are the hopping energy and 2D renormalized interaction, respectively. Terms proportional to $\hat{a}^{\dagger} \hat{a}$ and $\hat{b}^{\dagger} \hat{b}$ just count the total number of atoms in the system, and have been neglected in Eq.~(\ref{DiracHubbard}) as an overall constant. Equation~(\ref{DiracHubbard}) is the Hubbard Hamiltonian divided into two degenerate sublattices A and B, appropriate to the honeycomb optical lattice. 



The time evolution of the bosonic operators $\hat{a}$ and $\hat{b}$ is computed according to the standard Heisenberg prescription: $i\hbar \,\partial_t\hat{a} = [\hat{a} , \hat{H}]$. Although we suppress the lattice site indices, implicitly the operator $\hat{a}_{i , j}$ which destroys a boson at site $(i , k)$ satisfies the bosonic commutation relation $[\hat{a}_{i, j} ,\hat{a}^{\dagger}_{i', j'}] = \delta_{i i'}   \delta_{j j'}$. Using this fact and a similar relation for $\hat{b}_{i, j}$, we compute the commutator of $\hat{a}$ with the Hamiltonian Eq.~(\ref{DiracHubbard}) to obtain
\begin{eqnarray} 
\fl i\hbar \,\partial_t\hat{a}_{i, j}  =   - t_h \left[\hat{b}_{i, j} e^{i {\bf k}\cdot ({ \bf r}_{A_{i,j} }-{ \bf r}_{B_{i,j}}) } + \hat{b}_{(i ,j )- {\bf n}_1} e^{i {\bf k}\cdot ({ \bf r}_{A_{i,j} }- {\bf r}_{B_{(i,j)- {\bf n_1}}}) } + \hat{b}_{(i, j) - { \bf n}_2}\,e^{i {\bf k}\cdot ({\bf r}_{A_{i, j} }- { \bf r}_{B_{(i, j) - {\bf n}_2}}) }\right] \nonumber \\
+  \, U  \hat{a}^{\dagger}_{i, j}  \hat{a}_{i, j}  \hat{a}_{i, j} \label{eqn:heisenbergA}
\end{eqnarray}
where the first three terms on the right hand side represent transitions from the three B-sites nearest the $(i,j)^{\mathrm{th} }$ site of the A sublattice and ${\bf n}_1$ and ${\bf n}_2$ are primitive cell translation vectors for the reciprocal lattice, as shown in Fig.~\ref{Honeycomb_Lattice}(a). The same method yields a similar equation for the B sublattice. We next calculate the time rate of change of the expectation value of Eq.~(\ref{eqn:heisenbergA}) with respect to on-site coherent states. For the $(i, j)^\mathrm{th}$ site this is defined as 
\begin{eqnarray}
| a_{i,j} \rangle = e^{- |a_{i, j} |^2 /2} \sum_n a_{i,j}^n \frac{ ( \hat{a}_{i, j}^\dagger  )^n}{n !}   |0 \rangle \, ,  \label{coherentstate}
\end{eqnarray}
which is an eigenstate of the destruction operator $\hat{a}_{i , j}$ with eigenvalue $a_{i,j} \in \mathbb{C}$, as can readily be verified. Taking the full wavefunction to be the direct product over all $M \times M$ lattice sites (including the B sublattice), we have 
\begin{eqnarray}
| a , b ; M \times M \rangle \equiv  \bigotimes_{i, j; k, l }^{M \times M}   \left(  | a_{i,j} \rangle \bigotimes   | b_{k, l}  \rangle \right)  \, . \label{productcoherent}
\end{eqnarray}
From the definition of the single-site coherent state Eq.~(\ref{coherentstate}), one may then verify that 
\begin{eqnarray}
\hat{a}_{i,j}  | a , b ; M \times M \rangle  =  a_{i,j}  | a , b ; M \times M \rangle  \, , 
\end{eqnarray}
and 
\begin{eqnarray}
  \langle a , b ; M \times M |    \hat{a}_{i,j}^\dagger   =     \langle a , b ; M \times M |    a_{i,j}^*  \, . 
\end{eqnarray}
Inserting the nearest-neighbor vectors ${\bf \delta}_1$, ${\bf \delta}_2$, and ${\bf \delta}_3$ into the exponentials in Eq.~(\ref{eqn:heisenbergA}) (see Fig.~\ref{Honeycomb_Lattice}(b)), and doing the same in the equation for $\hat{b}$, then taking the expectation with respect to the state Eq.~(\ref{productcoherent}) we obtain coupled equations of motion for discrete, on-site, complex-valued amplitudes
\begin{eqnarray} 
\fl i\hbar \,\dot{a}_{i, j}  =   - t_h \left[  b_{i, j} e^{i {\bf k}\cdot {\bf \delta}_3 } +  b_{(i ,j )- {\bf n}_1} e^{i {\bf k}\cdot   {\bf \delta}_1    } +  b_{(i, j) - { \bf n}_2}\,e^{i {\bf k}\cdot  {\bf \delta}_2 }\right] +   U   a^{*}_{i, j}  a_{i, j}  a_{i, j} \label{eqn:heisenbergA}\, ,   \\
\fl i\hbar \,\dot{b}_{i, j}  =   - t_h \left[  a_{i, j} e^{ - i {\bf k}\cdot {\bf \delta}_3 } +  a_{(i ,j ) + {\bf n}_1} e^{ - i {\bf k}\cdot   {\bf \delta}_1    } +  a_{(i, j) + { \bf n}_2}\,e^{ - i {\bf k}\cdot  {\bf \delta}_2 }\right] +   U   b^{*}_{i, j}  b_{i, j}  b_{i, j} \label{eqn:heisenbergB}
\end{eqnarray}
The NLDE is derived around the linear band crossings between the A and B sublattices at the Brillouin zone corners~\cite{wallacePR1947}, called the \emph{Dirac cones} in the graphene literature~\cite{geim2007}. To this end, we insert appropriate values for the the nearest-neighbor displacement vectors ${\bf \delta}$ and evaluate the wave vector ${ \bf k}$ at the Brillouin zone corner, defined by ${ \bf k} = {\bf K} = (0,4\pi/3)$ , ${\bf \delta}_1=(1/2\sqrt{3} , - 1/2 )$, ${ \bf \delta}_2=( 1/2\sqrt{3} , 1/2 )$, ${\bf \delta}_3=(-  1/\sqrt{3} ,0)$. Finally, taking the long-wavelength limit of the resulting equations recovers a continuum theory but with a Weyl spinor wavefunction $\Psi = (\psi_A , \, \psi_B)$.

The key point in discerning the correct normalization (and thus other related quantities) is the contraction of the many-body bosonic operators between localized coherent states. The parameters $|a_{i,j}|^2$ and $|b_{i,j}|^2$ which label the coherent states for the A and B sublattices at site $(i, j)$, emerge as the number of atoms at each site, so that $a_{i,j}$ and $b_{i,j}$ become continuous amplitudes $\psi_A({\bf r}, t)$ and $\psi_B({\bf r},t)$ in the long-wavelength limit, as stated above. Note that the complex moduli of these amplitudes are pure dimensionless particle numbers, not densities, since they result from taking the spatial integral over the lattice. With the area per lattice site given by $A_l = \sqrt{3} a^2/4$, the local time-dependent sublattice densities can be reconstructed as: $\rho_{A (B)}({\bf r},t) = |\psi_{A(B)}({\bf r}, t)|^2/ A_l$. Then, the dimensionally correct sublattice mean-field wavefunctions must be given by $\psi_{A(B)}({\bf r}, t) / \sqrt{A_l} = (16/3 a^4)^{1/4} \, \psi_{A(B)}({\bf r}, t)$, where $a$ is the usual lattice spacing. The correct normalization procedure can now be deduced by writing down the total number of particles in the system
\begin{eqnarray}
N  =   (16/3 a^4)^{1/2} \!  \int_0^{2 \pi}  \! \! d \phi \int \!  r  \, dr ( |\psi_A(r, \phi; t)|^2 + |\psi_B(r,\phi; t)|^2)  \,  , \label{nor}
\end{eqnarray}
where the upper limit of the radial integral is taken large enough so that the integrand is negligible. The total number of atoms of the system, $N$, appears on the left-hand side.

The 3D to quasi-2D reduction and continuum regime result in an effective atomic interaction $U$, a renormalized version of the usual interaction $g$. We arrive at the explicit form for $U$ by first approximating the lowest band on-site Wannier functions by the ground state of the harmonic oscillator potential. Integrating over the area of one site, we obtain a new local interaction strength
\begin{eqnarray}
U  \equiv g_\mathrm{2D}\, \left( \! \frac{\sqrt{3} a^2}{4} \right)^2 \bar{n}_\mathrm{2D}^2  \int \! dx dy\,  |w_i(x,y)|^4  \nonumber \\
 =  g_\mathrm{2D}  \,  \left( \! \frac{\sqrt{3} a^2}{4} \right)^2 \bar{n}_\mathrm{2D}^2 \left(\frac{1}{2 \pi \ell^2} \right)\, , 
\end{eqnarray}
where $\ell$ is the oscillator length of a lattice potential well. It is often more practical to express the area of one site in terms of the lattice constant $\pi \ell^2 = \sqrt{3} a^2/4$, and all other parameters in terms of the corresponding 3D parameters. Using Eqs.~(\ref{g2D})-(\ref{nbar2D}), the interaction takes the form
\begin{eqnarray}
U =     L_z    \, g \, \bar{n}^2   \,  \frac{3  \sqrt{3}\,  a^2}{8} \,  . \label{effint}
\end{eqnarray} 
Note that $U$ has dimensions of energy.

We can now identify the main parameters which appear in the NLDE. The dimensionful coefficient which multiplies the Dirac kinetic term is the effective speed of light $c_l \approx 5.31 \times 10^{-2}\,  \mathrm{cm/s}$ (compare to the analogous coefficient for relativistic electrons $c \approx 3.00 \times 10^8 \, \mathrm{m/s}$). In terms of fundamental constants we find $c_l \equiv  t_h a \sqrt{3} /2 \hbar$, where $a$ is the lattice constant and $t_h$ is the hopping energy. The natural length scale of the NLDE is the \emph{Dirac healing length} $\xi_\mathrm{Dirac} \equiv  \hbar c_l/U  =  t_h a \sqrt{3} /2 U$, which characterizes the distance over which a disturbance of the condensate will return to its uniform value. We see that $\xi_\mathrm{Dirac}$ has the correct dimension of length. To simplify the notation, for the remainder of our paper we will omit the 2D subscript on all parameters with typical values as can be achieved in present experiments~\cite{Haddad2012}. Finally, the quantity $U$ which appears in the NLDE determines the strength of the nonlinearity. We have provided a full list of relevant parameters associated with the NLDE in Table~\ref{table1}. 
\begin{table*}[]
\resizebox{15.5cm}{!}{
\begin{tabular}{llll}
\multicolumn{4}{c}{}  \vspace{.5pc}\\
\hline
\hline 
Parameter &\hspace{1pc}  Symbol/Definition & Value \hspace{2.5pc} &  Range  \hspace{1pc} \vspace{.5pc}  \\
\hline
Plank's constant\; & \; $\hbar$  \;  &  $1.06 \times 10^{-34}\,  \mathrm{j}\cdot \mathrm{s}$ \; &    \\
Boltzman's constant    \;    &\;  $k_B$ \;   & $1.38 \times 10^{-23} \,  \mathrm{j} \cdot \mathrm{K}^{-1}$ \; &  \\
Mass of $ ^{87}\mathrm{Rb}$  & \; $M$  & $1.44  \times 10^{-25}\,  \mathrm{kg}$ \;  &  \\
Number of atoms & \; $N$ \; & $ 3.00 \times 10^{4}  $ \; & $10^2 - 10^{10}$  \\
Wave number of laser light   &\; $ k_L$ \;  & $ 7.57 \, \times 10^6\,  \mathrm{m}^{-1}$ \;  &  $4.19 \times 10^6-  4.19 \times 10^7\mathrm{m}^{-1}$ \\
Lattice constant & \; $a = 4 \pi/ 3k_L$ \; & $0.55  \, \mathrm{\mu m}$ \; &  $0.30- 0.70 \, \mathrm{\mu m}$ \\
Recoil energy  & \; $E_R = \hbar^2 k_L^2/2M$ \; & $0.16 \, \mathrm{\mu K}$   \;         &  $0.049  - 4.90\, \mathrm{\mu K}$ \\
Lattice potential & \; $V_0 = 16 E_R$ \; &   $10.1   \, \mathrm{\mu K}$ \;           &  $0.79 - 10.1 \, \mathrm{\mu K}$    \\
Hopping energy   &\; $t_h  = 1.861 \left(V_0/E_R\right)^{3/4} E_R\, e^{ -1.582 \sqrt{V_0/E_R} }$ \;  &$16.8 \,  \mathrm{nK}$ \; &  $3.49 \, \mathrm{nK} - 1.90  \, \mathrm{\mu K}$  \\
Scattering length  & \; $a_s$ \; & $5.77  \, \mathrm{nm}$ \; &   $5.00  - 10.0\, \mathrm{nm}$  \\
Average particle density   & \; $\bar{n}$ \; & $5.86 \times 10^{18} \, \mathrm{m}^{-3}$ \; &  $10^{15} - 10^{21} \, \mathrm{m}^{-3}$ \\
Two-body interaction  & \; $g = 4 \pi \hbar^2 a_s/M $ \; & $41.0    \, \mathrm{K}\cdot \mathrm{nm}^3$\; & $22.36 - 52.18  \, \mathrm{K}\cdot \mathrm{nm}^3$                      \\
Healing length  & \; $\xi = 1/\sqrt{8 \pi \bar{n} a_s} $ \; &  $1.10  \, \mathrm{\mu m}$ \; & $ \lesssim 1.50 \, \mathrm{\mu m}$  \\
Sound speed & \; $c_s = \sqrt{g \bar{n}/M} $ \; & $4.82 \times 10^{-2} \, \mathrm{cm}/\mathrm{s}$ \;  &  $5.83 \times 10^{-3} - 0.825 \, \mathrm{cm}/\mathrm{s}$  \\
Sound speed (2D) & \; $c_{s \mathrm{2D}} =   \left( 3/2 \right)^{1/2} \!  c_s$ \; & $5.90  \times 10^{-2} \,   \mathrm{cm}/\mathrm{s}$  &  $7.14 \times 10^{-3} - 1.01 \, \mathrm{cm}/\mathrm{s}$  \\
Healing length (2D) & \; $\xi_{\mathrm{2D}} =  \left(2/3\right)^{1/2}\!   \xi $ \; & $1.75  \, \mathrm{\mu m}$  &  $ \lesssim 2.45 \, \mathrm{\mu m}  $    \\
Transverse trap energy & \;  $\hbar  \omega_z$ \; &  $22.17  \, \mathrm{nK}$ \; &  $0.21  - 56.5 \, \mathrm{nK}$ \\
Transverse oscillator length  & \;  $L_z =( \hbar /M \omega_z)^{1/2} $ \; &  $1.50 \, \mathrm{\mu m}$ \;  & $< 3.0  \, \mathrm{\mu m}$ \\
Average particle density (2D)  & \; $\bar{n}_\mathrm{2D} =  L_z \, \bar{n}  $ \; & $4.50 \times 10^{12} \, \mathrm{m}^{-2}$ \; & $ 10^{9} -  5.00 \times 10^{15} \, \mathrm{m}^{-2}  $  \\
Effective speed of light  & \;  $c_l = t_h a \sqrt{3}/2 \hbar$ \; &  $5.31  \times 10^{-2}\, \mathrm{cm}/\mathrm{s} $ \; & $ < 5.40 \times 10^{-2}  \, \mathrm{cm}/\mathrm{s}$ \\
Dirac kinetic coefficient & \;  $\bar{c}_l  = \hbar c_l$ \; &  $2.07 \, \mathrm{nK} \cdot \mathrm{\mu m}$ \; & $ < 5.72 \,  \mathrm{nK} \cdot \mathrm{\mu m}$  \\
Dirac nonlinearity  & \;  $U =     L_z    \, g \, \bar{n}^2   \, 3  \sqrt{3}\,  a^2 /8$   &  $1.07 \, \mathrm{nK}$  &  $ < 2.36 \, \mathrm{nK}$  \\
Dirac healing length  & \;  $\xi_\mathrm{Dirac} = t_h a \sqrt{3}/2 U$ \; & $3.80\, \mathrm{\mu m}$ & $0.50 - 50.0 \, \mathrm{\mu m}$   \vspace{.5pc} \\
\hline
\hline
\end{tabular}}
\caption{Physical Parameters for the NLDE typical for a BEC of $^{87}$Rb atoms. The renormalized parameters are expressed in terms of fundamental quantities. The range of possible values account for the physical constraints discussed in the main text.} \label{table1}
\end{table*}

\subsection{Physical constraints}

The realization of the NLDE in a condensate of $^{87}$Rb atoms requires that several constraints are satisfied which we now list and discuss:

\begin{enumerate}

\item  \emph{Landau Criterion}.~In order to avoid the instabilities associated with propagation faster than the sound speed in the condensate, we require that the effective speed of light is less than the 2D renormalized speed of sound. 

\item \emph{Long-wavelength Limit}.~The NLDE describes propagation of the long-wavelength Bloch envelope of a BEC near the Dirac point. Thus, a necessary condition for realizing the NLDE in the laboratory is that the Dirac healing length must be much larger than the lattice constant. 

\item \emph{Relative Lengths for 2D Theory}.~In order to obtain an effectively 2D system, the vertical oscillator length must be much smaller than the trap size along the direction of the plane of the condensate. 

\item \emph{Relative Energies for 2D Theory}.~Analogous to the previous restriction, this condition relates to the 2D structure but pertains to the energies of the system. The key point is that we must avoid excitations vertical to the plane of the condensate while enabling them along the plane: the chemical potential and temperature must be less than the lowest transverse excitation energy.

\item \emph{Weakly Interacting Regime}.~The NLDE and RLSE are derived for a weakly interacting Bose gas. This ensures both the stability of the condensate as well as the effective nonlinear Dirac mean-field description. We then require the interaction energy to be significantly less than the total energy of the system.

\item \emph{Dirac Cone Approximation}.~For a condensate in the regime where the NLDE description is valid, we require that the linear approximation to the exact dispersion remain valid. As in the case of graphene, large deviations from the Dirac point induce second order curvature corrections to the dispersion. Thus, we must quantify the parameter restrictions which allow for a quasi-relativistic interpretation.$^{\footnotemark[2]}$ \footnotetext[2]{We note that the Dirac cone approximation is not necessarily adhered to in analogous honeycomb photonic lattice systems. See for example Refs.~\cite{Peleg2007,Bahat2008}.  }

\item \emph{Lowest Band Approximation}.~We derive the NLDE and RLSE assuming that the lowest band is the main contribution to the dispersion.

\end{enumerate}

Having stated each constraint, we can now address each one in detail and explore the conditions under which each is satisfied. In the following, we consider a BEC comprised of $^{87}\mathrm{Rb}$ atoms where all numbers used are listed in Table~\ref{table1} and are experimentally realistic~\cite{Ohara2005}. First, the Landau criterion pertains to the effective velocities in the BEC. Stated mathematically, the Landau criterion requires that $c_l / c_{s  \mathrm{2D}}  <1$. Using the definitions for the effective speed of light and the sound speed found in the first part of this section, we compute $c_l / c_{s \mathrm{2D}}  =  0.90$, which satisfies the inequality.

The length constraints are as follows. The long-wavelength limit is defined by $\xi_\mathrm{Dirac} /a \gg 1$, for which we find that $\xi_\mathrm{Dirac} /a = 6.91$. For an effectively 2D system, the required length constraint implies the condition $L_z \ll R$. Taking $R \approx 100\,  a$ (a typical condensate size), and using a realistic value for the vertical oscillator length (Table~\ref{table1}), we obtain $L_z = 2.73 \, a$, which satisfies the constraint. Moreover, we require a healing length close to or less than the transverse oscillator length. With $\xi = 1.10 \, \mathrm{\mu m}$ and $L_z = 1.50 \,  \mathrm{\mu m}$, we find that this condition holds.

The energy constraints may be stated as $\mu, k_B T \ll \hbar \omega_z$. We can solve the NLDE for the lowest excitation to obtain an expression for the chemical potential $\mu = \hbar c_l k + U |\Psi|^2$~\cite{haddad2011}. Next, we evaluate this expression using the lowest excitation in a planar condensate of radius $R \approx 100 a$, which has wavenumber $k \approx \pi/2R= 2.86 \times 10^4 \, \mathrm{m}^{-1}$. The interaction $U$ is computed using Eq.~(\ref{effint}) for the binary interaction $g$  and mass $M$ pertaining to a condensate of $^{87}\mathrm{Rb}$ atoms. Finally, for a uniform condensate we take $|\Psi|^2 \approx 4/\sqrt{3} $ (Eq.~(\ref{nor})) and the constraint on the chemical potential becomes $\mu = 2.59 \, \mathrm{nK} < 22.17\, \mathrm{nK}$, which is satisfied. For the temperature, we require $ T\ll \hbar \omega_z /k_B$. Using the data in Table~\ref{table1} for the vertical oscillator frequency, we obtain the upper bound for the temperature $T \ll  22.17 \,  \mathrm{nK}$. This is a reasonable requirement given that BEC occurs for $T$ in tens or hundreds of nanoKelvins or as low as picoKelvins.

Next, we examine constraints on the particle interaction. To check that we are in the weakly interacting regime, i.e., that $U/\mu \ll 1$, we use the value for the chemical potential $\mu$ which we have just computed and compare this to the interaction energy $U$, whereby we find that $U/\mu = 0.41$. An essential feature of the NLDE is that characteristic fluctuations are close enough to the Dirac point so that the linear Dirac cone approximation remains valid. Expanding the exact dispersion near the Dirac point, we obtain $\mu(k) =  U \pm t_h \left(  a \sqrt{3}k /2+ a^2 k^2/8 - a^3 \sqrt{3} k^3/48 + ...   \right)$, where $k$ is a small deviation away from the Dirac point. The first term gives the linear Dirac dispersion. Higher order corrections describe curvature of the band structure away from the Dirac point. From the second order term we see that the NLDE description is valid for $a k /\sqrt{8} \ll 1$. This determines a lower bound on the wavelength for fluctuations of the condensate: $\lambda_\mathrm{min} \gg (2 \pi/\sqrt{8}) a$. Linear dispersion places an additional constraint on the chemical potential: $ |\mu| \ll  U + 6 t_h \simeq 101.9 \, \mathrm{nK}$. From the value of the chemical potential already obtained, we find $\mu = 2.59 \,  \mathrm{nK}  \ll  101.9 \, \mathrm{nK}$. Finally, weak short range interactions at very low temperatures justifies a lowest-band approximation to describe the physics of the NLDE.

 \section{Superfluid excitations near a Dirac point} 
 \label{Excitations}
 
The mean-field physics of single-particle states for a collection of fermions with Fermi energy near a Dirac point of a honeycomb lattice has been studied exhaustively and is discussed in various comprehensive articles~\cite{wuCongjun2008,Lee:2009,Geim:2007,Bittner2010}. For systems of bosons, however, one must carefully consider the meaning of condensation in the presence of Dirac points. To discuss BECs and Dirac points together one must address the compatibility of single-valuedness for phase functions required for stable vortex formation in a proper superfluid description, with the half-angle phase winding when circumnavigating a single Dirac point, i.e., the geometric or Berry phase~\cite{Nakahara2003}.

\subsection{Geometric and dynamical phase structure}

 \begin{figure}
\centering
\hspace{0pc} \includegraphics[width=.8\textwidth]{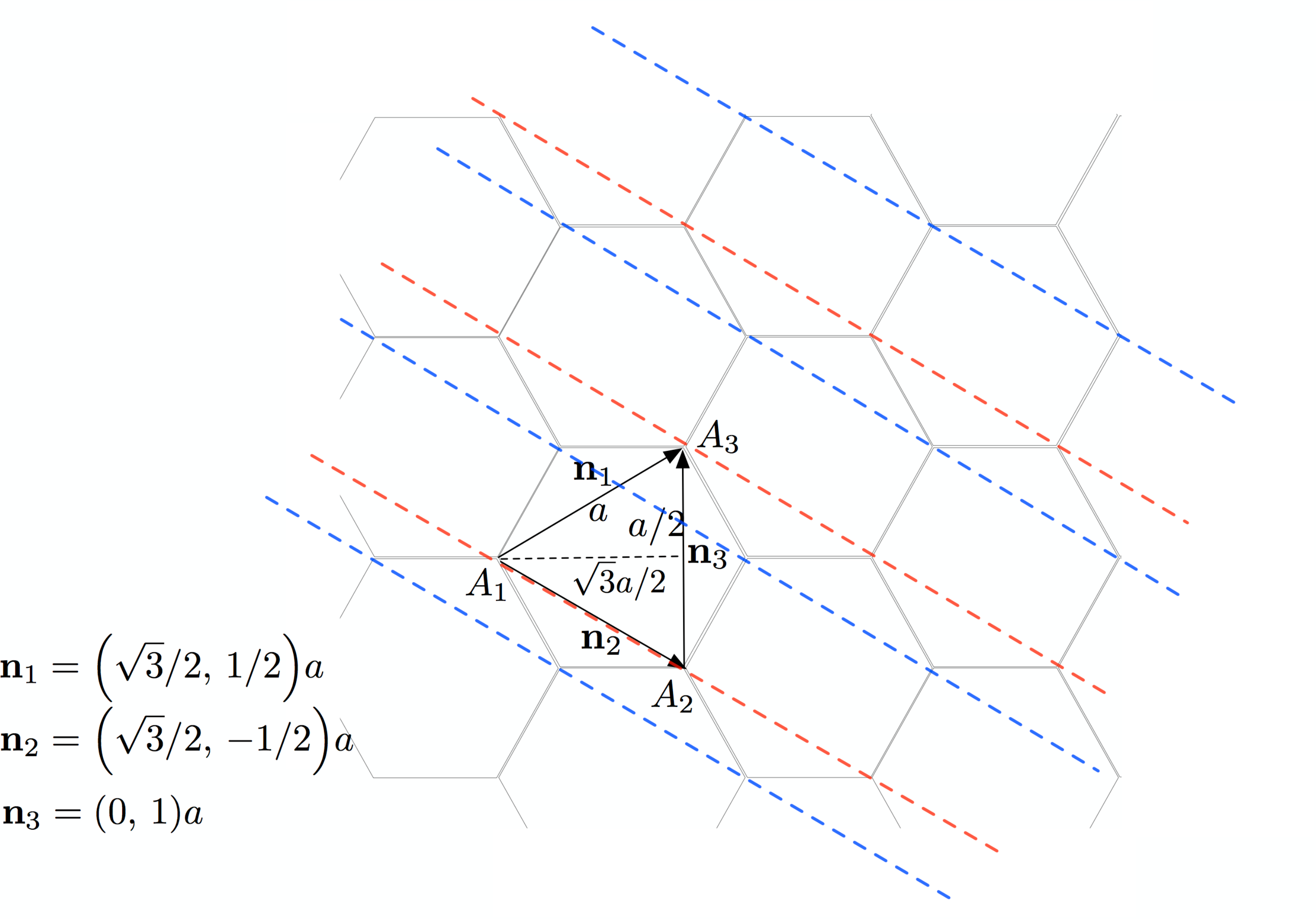} 
\hspace{0pc} \caption{(color online)     \emph{Bragg wave structure at the Dirac point of the honeycomb lattice}. Constructive interference of Bragg reflected waves for the Dirac point wavevector ${\bf K}$ (see Fig.~\ref{Honeycomb_Lattice}(c)) produces wavefront density peaks at A (red) and B (blue) sublattices. } 
\label{lattice}
\end{figure}

To address these issues, we first review some relevant information treated in most review articles on graphene, as this information is true for cold bosonic atoms as well~\cite{haddad2009}. The single-particle spectrum of the honeycomb lattice exhibits zero-points, or Dirac points, in the reciprocal lattice associated with crystal momentum ${\bf K} = ( 0, \, \pm 4\pi/3 a)$ rotated by $0, 2 \pi/3 , \, 4 \pi/3$, where $a$ is the lattice constant shown in Fig.~\ref{lattice}. Dirac points occur when the crystal momentum is tuned to the natural periodicity of the lattice with standing waves established due to Bragg scattering of the wave function. Reflection at the Brillouin zone edge is shown in Fig.~\ref{lattice}, where one adds up projections of the vectors ${\bf n}_1$ and ${\bf n}_3$ along the direction of the crystal momentum vector ${\bf K}$ connecting points on the A sublattice of equal phase, $A_1$ and $A_2$, to a third point $A_3$. In particular, at the Dirac point this sum results in a net $2 \pi$ accumulated phase angle at $A_3$. In Fig.~\ref{lattice}, the A and B sublattice wavefront density peaks are shown as red and blue dashed lines, respectively. In the tight-binding limit, the full lattice Hamiltonian reduces to two operators which couple the degenerate triangular A and B sublattices. The single-particle dispersion is computed by solving the $2 \times 2$ eigenvalue problem in momentum space determined by the Hamiltonian 
 \begin{eqnarray}
 \hat{H}({\bf k}) =    \left( \begin{array}{c c }
                                0 &     E({\bf k})^*  \\
                                    E({\bf k})    &  0  \\
                                \end{array} \right)  \, ,    \label{HamMatrix}
 \end{eqnarray}
 where the matrix elements come from computing the sublattice hopping energies
 \begin{eqnarray}
 E({\bf k}) =  - t_h  \left(  1 + e^{ i  {\bf n}_1 \cdot {\bf k}     } + e^{i   {\bf n}_2 \cdot  {\bf k} }  \right)  =  | E({\bf k}) | e^{- i  \phi({\bf k})}  \, .  \label{EofK}
 \end{eqnarray}
 Specifically, one finds
 \begin{eqnarray}
 |E({\bf k})| = t_h \sqrt{ 1 + 4 \,  \mathrm{cos} \! \left( \! \frac{\sqrt{3} k_x a}{2}\right)     \mathrm{cos} \! \left( \! \frac{ k_y a}{2}\right)  + 4 \,  \mathrm{cos}^2 \! \left( \! \frac{ k_y a}{2}\right)} \, . 
 \end{eqnarray}
 The eigenfunctions of $\hat{H}$ in Eq.~(\ref{HamMatrix}) are 
 \begin{eqnarray}
  \Psi_\pm({\bf k}) =    \left( \begin{array}{c }
                                     e^{ i \ell    \phi({\bf k}) }    \\
                                 \pm   \;     e^{ i ( \ell + 1)  \phi({\bf k}) }       \\
                                \end{array} \right)    \, ,  \label{DegenerateStates}
 \end{eqnarray}
 with eigenvalues $\pm | E({\bf k}) |$. Physically, the parameter $\ell \in \mathbb{R}$ comes from an extra U(1) phase degeneracy and reflects the gapless symmetry of the system under spatial translations of the atomic density at the Dirac point. The matrix in Eq.~(\ref{HamMatrix}) describes the amplitude and phase associated with real-particle tunneling between neighboring lattice sites. In particular, the phase of the wavefunction gets multiplied alternately by factors of $e^{\pm  i  \phi({\bf k})}$, so that no net phase is accrued when circumnavigating a closed path in the lattice. In contrast, long wavelength modes propagating in the lattice are described by linearizing the phase angle $\phi({\bf k})$ so that the local lattice scale variations in the phase structure are neglected, in which case one should expect a net phase accumulation.

We are particularly interested in this net geometric or Berry phase since we must factor it into the phase winding for vortex solutions of the NLDE. Although most treatments of the subject use a momentum space argument, here we use instead a more direct analysis in real space. We expand the Hamiltonian and eigenstates near the Dirac point by taking ${\bf k} = {\bf K} + {\bf  \delta k}$, with ${\bf K} = (4 \pi/3a ) { \bf \hat{y}}$ and ${ \bf \delta k}$ the small expansion parameter, i.e., we consider small deviations from the sublattice Brillouin zone corner. In real space, this amounts to a derivative expansion of Eq.~(\ref{EofK}) in terms of the directional derivatives ${\bf n}_1 \cdot {\bf \nabla}$ and ${\bf n}_2 \cdot {\bf \nabla}$. The first-order term gives the massless Dirac Hamiltonian and Dirac equation~\cite{haddad2009}, while higher products of derivatives provide corrections that probe the finer details of Bragg scattering around the Dirac point. To isolate the geometric phase, we consider adiabatic transport around a closed loop. Adiabaticity ensures that we do not accumulate a dynamical contribution to the phase and restricts the path to energy eigenstates states nearest the Dirac point. A direct way to accomplish this is by linearizing Eq.~(\ref{HamMatrix}) in real space, solving for the eigenstates in plane-polar coordinates $r$ and $\theta$, and restricting to paths with large radii $R \equiv r \gg a $. The large radius limit allows us to access only the longest wavelength modes that vary mainly tangentially with minimal radial contribution. Equation~(\ref{HamMatrix}) then reduces to
\begin{eqnarray}
\hat{H}(\theta) \approx  \frac{i  \hbar c_l }{ R} \left( \begin{array}{c c }
                                0 &     -  e^{-i \theta} \partial_\theta   \\
                                      e^{ i \theta} \partial_\theta    &  0  \\
                                \end{array} \right)  \, ,    \label{RotHam}
\end{eqnarray}
from which we find the eigenstates
\begin{eqnarray}
  \Psi_\pm(\theta) =    \left( \begin{array}{c }
                                     e^{ - i  \theta /2  }    \\
                                 \pm   \;     e^{ i  \theta/2 }       \\
                                \end{array} \right)    \, ,      \label{halfstates}
\end{eqnarray}
and energies $\pm \, \hbar  \omega/2$, with $\omega \equiv c_l /R$ and $c_l$ the effective speed of light. Note that in this limit the degeneracy in Eq.~(\ref{DegenerateStates}) is lifted and eigenstates are forced into the form Eq.~(\ref{halfstates}), which acquires a net phase of $\pi$ under a full $2 \pi$ rotation. Thus, linearization of Eq.~(\ref{HamMatrix}) leads to a double wrapping of the phase angle $\phi$ around the polar angle $\theta$. 
 
 In the case of a vortex, we can compensate for the Berry phase by requiring half-winding in the overall dynamical phase multiplying the spinor order parameter: $\exp[i(\ell + 1/2) \theta]$. As a result, the geometric phase becomes identified with the relative phase between the two sublattices. Hence, stable vortices are required to have half-integer internal geometric winding plus an overall half-integer dynamical winding such that the superfluid velocity is the sum of the gradients of both phases, effectively splicing together the internal and external phase.

\subsection{Superfluid regime and dimensional analysis}

To address superfluidity in the honeycomb lattice using a semiclassical approach, consider a thermal excitation with crystal momentum ${\bf p}$ (measured from the Dirac point) interacting with an atomic gas at the Dirac point (${\bf p}_\mathrm{Dirac} \equiv 0$) producing an excitation in the gas with momentum ${\bf p}^\prime$. It follows from energy conservation, $\Delta E  =0$, that
\begin{eqnarray}
\Delta E  =  \pm \, \frac{ \left| {\bf p} - {\bf p}^\prime \right|^2 }{2 m^*} \mp c_l | {\bf p} - {\bf p}^\prime | + E({\bf p}^\prime )  \mp \frac{ \left| {\bf p}  \right|^2 }{2 m^*} \pm c_l | {\bf p} | = 0 \, .  \label{firstcondition} 
\end{eqnarray}
 In Eq.~(\ref{firstcondition}), for generality we assume that the incoming thermal mode may be in a Bloch state far enough removed from the Dirac point such that second-order corrections are important. Thus, $m^*$ is the effective mass related to the dispersion curvature, $E({\bf p}^\prime )$ is the energy of the quasiparticle excitation in the gas, and the upper and lower signs refer to negative and positive dispersion branches, respectively. We first examine the linear regime for which $p \approx p^\prime << c_l m^*$, in which case we can neglect quadratic terms in Eq.~(\ref{firstcondition}). Keeping only linear terms and using $ | {\bf p} - {\bf p}^\prime | = \sqrt{ p^2 - 2 p p^\prime \mathrm{cos} \theta + {p^\prime}^2 }$, with $p = |{\bf p}|$ and $\theta$ the angle between ${\bf p}$ and ${\bf p}^\prime$, Eq.~(\ref{firstcondition}) forces the constraint 
 \begin{eqnarray}
\mathrm{cos} \theta = \pm \,  \frac{E({\bf p}^\prime)}{ c_l p^\prime }  \, , 
\end{eqnarray}
and four conditions determined by the different sign combinations for incoming and scattered modes. When the signs of the incoming thermal and scattered condensate modes are the same we find that $\theta = 0$. On the other hand, if the energies of incoming and scattered modes have opposite sign, we obtain $\theta = \pi$, thus scattering in the reverse direction occurs between Dirac particles and anti-particles as one should expect. Notice that in this linear regime conservation of energy places no additional constraints on $p$ and $p^\prime$, so that in our mean-field analysis an equilibrium between condensate and non condensate atoms is maintained: the incoming mode transfers all of its energy and momentum to an excitation of the condensate leaving a single outgoing excitation with the same energy and momentum.

 A second regime of Eq.~(\ref{firstcondition}) corresponds to the condition $p^\prime <<   p  <  c_l m^*$, which yields the constraint 
 \begin{eqnarray}
p \,  \mathrm{cos} \theta   = \pm \, m^* \frac{E({\bf p}^\prime)}{p^\prime} \; , \label{secondcondition}
\end{eqnarray} 
leading to the same conditions as in the previous case for the scattering angle $\theta$, but now with an additional constraint for an upper bound critical momentum $p_c$ below which no excitation can be created in the condensate
\begin{eqnarray}
p_c  = m^* \, \mathrm{min}_{{\bf p}^\prime} \frac{| E({\bf p}^\prime) | }{ p^\prime}  \, . \label{criticalmom}
 \end{eqnarray}
 Equation~(\ref{criticalmom}) recovers Landau's criterion for superfluidity but here in terms of the absolute value of the quasiparticle energy to account for scattering into negative energy states. We point out that the absolute value in Eq.~(\ref{criticalmom}) is a strictly a consequence of energy conservation and the presence of quadratic terms in Eq.~(\ref{firstcondition}); the various sign combinations in Eq.~(\ref{secondcondition}) are taken into account through the scattering angle $\theta$. With $E({\bf p}^\prime) = \pm c_l p$, the upper critical bound is just $p_c  = m^* c_l$. Below this value (and for $p >> p^\prime$) thermal modes cannot interact with the condensate, thus superfluidity is preserved. For $p > p^\prime$, however, as expected we see a breakdown of superfluity. In our analysis we have nowhere included details of the interaction; only a knowledge of states near the Dirac point was needed. Once we consider quantum effects and details of the interaction our results will change significantly, as we will see in Sec.~\ref{Stability}.


Consider again a non-condensate excitation with initial momentum ${\bf p}$ interacting with the condensate by transferring all of its momentum and energy to the condensate. A secondary excitation is then emitted with exactly the same momentum and energy. Since the initial and final excitations are indistinguishable, we can view this process as a single excitation interacting weakly with the condensate and continuing on its way with only an average self-energy correction. At the quantum level and to first order in the interaction $U$, a single interaction point must be averaged over the volume (area in quasi-2D) of the condensate. Since we are dealing with very long wavelengths the result is a nonlocal collective excitation formed as a composite of the initial incoming mode dressed by the condensate background. At long wavelengths linear perturbation couples particles and holes, which amounts to reducing the power in the dispersion relation $E(k)$ by a factor of $1/2$. In contrast, at shorter wavelengths (higher energies) the incoming excitation couples with a quasiparticle locally, so that the available states for thermal and condensate modes remain distinct. Elaborating further, the number of accessible states less than $k$ for the undressed excitation plus condensate is $\Omega(k) = a_r k^r$, and the dispersion is $E  = \pm c_l \hbar k^s $. Here for our argument we leave the constants $r$, $s$, and $a_r$ general. Thus, $\Omega(E) = 2 a_r \, E^{r/s}/(\hbar c_l)^{r/s}$ where the extra factor of $2$ accounts for both positive and negative eigenvalues. This yields the density of states 
\begin{eqnarray}
D(E) = \frac{d}{d E}  \Omega(E) = 2 a_r  \frac{ r}{s}  \frac{ E^{(r/s) -1 }}{ (\hbar c_l)^{r/s} } \, . 
\end{eqnarray} 
In order to maintain $D(E)$ constant when transitioning from short to long wavelengths, $c_l p  > U \to c_l p  <  U$, imposing the $1/2$-power reduction in the dispersion, i.e., $\Omega(k) = a_r k^r \to a_r k^{r/2}$, requires that we also take $s \to s/2$. The renormalized energy is then $E \propto \pm k^{1/2}$ (for $s =1$). The proportionality constant must involve the quasi-2D interaction $U$ which we determine through dimensional analysis to be
\begin{eqnarray}
E(p) = \pm \sqrt{ U c_l p  } \, . \label{lowenergyspec}
\end{eqnarray}

Note that Eq.~(\ref{lowenergyspec}) leaves out the possible form $E(p) = \pm \sqrt{ - U c_l p  }$, which displays a low-momentum dynamical instability. However, this is regularized by accounting for a finite system size which imposes a lower momentum cutoff $\vert {\bf p}\vert_\mathrm{min} = 2 \pi  \hbar /R$, where $R$ is the radial size of the condensate. For the usual harmonic trap with frequency $\Omega$ we have $R =  (\hbar/M \Omega )^{1/2}$. By dimensional analysis, in terms of the quasi-2D renormalized average particle density $\bar{n}$ and interaction $U$, we obtain the stability requirement for the oscillator length $R \le  \sqrt{c_l\pi \hbar /(\bar{n} U)}$. From a practical standpoint the lower bound $\vert {\bf p}\vert_\mathrm{min}$ removes the longest wavelength modes which opens an insulating buffer between the positive and negative parts of the spectrum in addition to regulating the dynamical instability.

\section{Relativistic linear stability equations}
\label{RLSE}

Bogoliubov's method was originally introduced in his 1947 paper~\cite{Bogoliubov1947} (see also~\cite{ Zagrebnov2001,Adams2004} for thorough contemporary treatments), and the concept later generalized by Fetter~\cite{Fetter1972} to accommodate nonuniform condensate profiles. The latter formulation gives a convenient method for computing quasiparticle states and the associated eigenvalues by substituting the spatial functions for a particular background condensate into a pair of coupled differential equations, and then solving the resulting eigenvalue problem. Fetter's extended method was designed with a vortex profile in mind, and has proven successful for computing stability of vortices in trapped condensates, but also for gaining a deeper understanding of general vortex dynamics~\cite{Fetter2001,Fetter2000,feder2000}. The set of equations that we derive in this section form the counterpart to Fetter's equations, but for trapped condensates that exhibit Dirac points in their dispersion~\cite{haddad2011}. We call them \emph{relativistic linear stability equations} because of the quasi-relativistic context here and the similarity to equations that appear in relativistic fluid dynamics. It is noteworthy that our result is not limited to the honeycomb optical lattice but applies generically to any system where the linear dispersion and bipartite structure are present, and where the contact interaction between constituent bosons is weak.

%

Our derivation of the RLSE relies fundamentally on Bogoliubov's method~\cite{Bogoliubov1947} as the underlying principle, and refers to Fetter's work~\cite{Fetter1972} for technical considerations regarding nonuniform condensates. First, we recall the second-quantized many-body Hamiltonian for weakly interacting bosons
 \begin{eqnarray}
 \hat{H}  =  \int \! d^2r  \, \hat{\psi}^\dagger H_0 \hat{\psi} + \frac{g}{2} \int \! d^2r\,  \hat{\psi}^\dagger\hat{\psi}^\dagger  \hat{\psi} \hat{\psi}\,,\label{eqn:H}\\
 \end{eqnarray}
where
\begin{eqnarray}
H_0 \equiv -\frac{\hbar^2}{2m}\nabla^2 + V({ \bf r})\,. 
 \end{eqnarray}
Here, $V({ \bf r})$ is the lattice potential and $g$ is the strength of the contact interaction. The first step is to decompose the wavefunction as the sum $\hat{\psi}({ \bf r})= \zeta({ \bf r}) \, \hat{a}_0 \,+\, \hat{\phi}({ \bf r})\label{eqn:psiBog}$, where we have split the wavefunction into a part that describes the condensate (first term) and satisfies the bosonic commutation relation $[ \hat{a}_0\,, \, \hat{a}_0^\dagger ]  \,=\, 1$, and a second part that describes small quasiparticle fluctuations. The operator in the first term destroys a particle in the mean-field $\zeta$, which, by itself, is a good approximation to $\hat{\psi}$. The second term destroys a particle in a number of single particle basis states of the noninteracting system, and describes the part of $\hat{\psi}$ that deviates from the mean field. Taking the Bogoliubov limit requires $\hat{a}_0 \, \rightarrow \,N^{1/2}_0$, where $N_0$ is the total number of condensed atoms, but we choose to compute the commutator before taking this limit in order to retain the effect of the presence of a macroscopic condensate field. We can obtain the commutation relations for $\hat{\phi}$ and $\hat{\phi}^\dagger$ by knowing that $\hat{\psi}$, $\hat{\psi}^\dagger$, and $\hat{a}_0$ and $\hat{a}^\dagger_0$ obey bosonic commutation relations. We obtain the quasiparticle commutation relations: $\left[ \hat{\phi}({ \bf r})\, , \,  \hat{\phi}^\dagger({ \bf r}')\right] =  \delta ({ \bf r} \,-\, { \bf r}')\, -\, \zeta({ \bf r}) \,  \zeta^*({\bf r}')$, $\left[ \hat{\phi}({ \bf r})\, , \,  \hat{\phi}({\bf r}')\right] = 0$, $\left[ \hat{\phi}^\dagger({ \bf r})\, , \,  \hat{\phi}^\dagger({\bf r}')\right] = 0$. In the Bogoliubov limit the condensate wavefunction has no operator part, in which case $\hat{\psi}$ may be written as $\hat{\psi}({ \bf r}) \,=\, \Psi({ \bf r}) \, +\, \hat{\phi}({\bf r})$. The condensate wavefunction has well defined phase and particle density and so may be expressed as: $\Psi({ \bf r}) = \sqrt{N_0/A} \,  e^{i S({ \bf r})} \sqrt{\rho({ \bf r})}$, where $A$ is the area covered by the planar condensate. Note that the amplitude is normalized as $A^{-1} \int \! d^2r \,\rho({r}) = 1$. With these definitions, the usual bosonic commutation relations become: $\left[\hat{\phi}({\bf r}) , \; \hat{\phi}^\dagger( \bf {r}')\right] =e^{iS({ \bf r})} \; e^{- iS({ \bf r}')} \; \bar{\delta}({ \bf r}, \; {\bf r}')$, where $\bar{\delta}({ \bf r}, {\bf r}')= \delta({ \bf r} - { \bf r}  ' ) -A^{-1} \sqrt{\rho( {\bf r})} \sqrt{\rho({ \bf r} ' )}$.

Next, we transform to the new Hamiltonian defined by  $\hat{K} = \hat{H} -  \mu \hat{N}  = \hat{H} - \mu \int \! d^2r \, \hat{\psi}^\dagger\, \hat{\psi}$, then expand through second order in the operator part eliminating the linear terms by forcing the condensate wavefunction to satisfy the constraint $( H_0 \,  -\, \mu \, +\, g \left|\Psi \right|^2 ) \Psi = 0$. We arrive at the Bogoliubov Hamiltonian $\hat{K}= \hat{K}_0 + \hat{K}_2$, wherein zero-order and second-order operator terms are grouped into $\hat{K}_0$ and $\hat{K}_2$ respectively. These are defined as
\begin{eqnarray}
\hat{K}_0  =  \int \! d^2r \, \Psi^*({ \bf r}) \left[ H_0 -\mu + \frac{g}{2}  \left| \Psi({ \bf r}) \right|^2\right] \Psi({ \bf r}) \, , \nonumber \\
\hat{K}_2  = \int \! d^2r \,  \hat{\phi}^\dagger({ \bf r}) \left[ H_0 -\mu  + 2 g  \left| \Psi({ \bf r}) \right|^2\right] \hat{\phi}({ \bf r}) \nonumber \\
 +  \frac{g}{2}  \int \! d^2r \, \left\{ \left[ \Psi^*({ \bf r})\right]^2\hat{\phi}({ \bf r}) \hat{\phi}({ \bf r})+\hat{\phi}^\dagger({ \bf r}) \hat{\phi}^\dagger({ \bf r})     \left[ \Psi({ \bf r})\right]^2  \right\}. \label{eqn:KBog2}
\end{eqnarray}
Note that in addition to the kinetic operator we also have an arbitrary external potential in the first two terms, which in our case will be the periodic potential of the optical lattice. Equation~(\ref{eqn:KBog2}) is quadratic in the field operators and so may be diagonalized with the appropriate field redefinition. To diagonalize Eq.(\ref{eqn:KBog2}) we first apply the linear transformation $\hat{\phi}({ \bf r}) = e^{i S({ \bf r})} \sum_j^{\prime} \left[ u_j({ \bf r}) \; \hat{\alpha}_j \; -\;v_j^*({ \bf r}) \; \hat{\alpha}^\dagger_j    \right]$ and $\hat{\phi}^\dagger({ \bf r}) =  e^{- iS({ \bf r})}   \sum_j^{\prime}   \left[ u_j^*({ \bf r}) \; \hat{\alpha}^\dagger_j \; -\;v_j({ \bf r}) \; \hat{\alpha}_j             \right]$, where the prime notation on the summation sign indicates that we are omitting the condensate from the sum. The $ \hat{\alpha}_j$'s and $ \hat{\alpha}_j^\dagger$'s inherit standard bosonic commutation relations from $\hat{\phi}$ and $\hat{\phi}^\dagger$, and the spatially dependent transformation coefficients $u_j({ \bf r})$ and $v_j({ \bf r})$ obey the completeness relations
\begin{eqnarray}
 \sum_j^{\;\;\;\;\;\prime} \left[ u_j({ \bf r}) \; u_j^*({ \bf r} ') \; -\;   v_j^*({ \bf r}) \; v_j({ \bf r} ')   \right]  =  \bar{\delta}({ \bf r} ,{ \bf r} ') \label{eqn:CompRel1}\\
\sum_j^{\;\;\;\;\;\prime} \left[ u_j({ \bf r}) \; v_j^*({ \bf r} ') \; -\;   v_j^*({ \bf r}) \; u_j({ \bf r} ')   \right]  =  0  \label{eqn:CompRel2}\\
 \sum_j^{\;\;\;\;\;\prime} \left[ u_j^*({ \bf r}) \; v_j({ \bf r} ') \; -\;   v_j({ \bf r}) \; u_j^*({ \bf r} ')   \right]  =  0   \label{eqn:CompRel3}\,  .
\end{eqnarray}

So far, our discussion has taken place in two continuous spatial dimensions constrained only at the boundary by a trapping potential. We now want to translate to a formalism that fits a two-dimensional periodic optical lattice potential with honeycomb geometry. This is done by assuming a tight-binding limit at each lattice site. Formally, this corresponds to expanding the wavefunction in terms of a Wannier basis: functions which are localized and centered on each lattice site. The nearest-neighbor approximation then allows for a decomposition of the condensate and operator parts in terms of individual sublattices labeled $A$ and $B$. In this new basis, the spatial dependence of the condensate and quasiparticle functions follows
\begin{eqnarray}
 \fl \Psi({ \bf r})  = \sum_A\,  e^{i { \bf k}\cdot ( { \bf r}- { \bf r}_A) }    \sqrt{n_{A_i}} \, e^{i S_{A_i} }       \,w( { \bf r}- {\bf r}_A)  + \sum_B  \, e^{i { \bf k}\cdot ( { \bf r}- { \bf r}_B) }  \sqrt{n_{B_i}} \, e^{i S_{B_i} } \,w( { \bf r}- { \bf r}_B) \, ,  \label{eqn:PsiBw2}   \\
 \fl \hat{\phi}( { \bf r}) = e^{iS( { \bf r })} \sum_{A,j}^{\;\; \;\prime}        \left[ u_{j,A_i}( { \bf r} -  { \bf r}_A) \hat{ \alpha}_j - v^*_{j,A_i}( { \bf r} -{ \bf r}_A)  \hat{\alpha}_j^\dagger \right]   \nonumber \\
 \fl +  \, e^{iS( {\bf r})}   \sum_{B,j}^{\;\; \;\prime}   \left[ u_{j,B_i}( { \bf r} -  {\bf r}_B) \, \hat{\beta}_j -  v^*_{j,B_i}( { \bf r} -  { \bf r}_B)  \hat{\beta}_j^\dagger  \right]     . 
 \label{eqn:SplitTrans1} 
   \end{eqnarray}

 \subsection{First method: Tight-binding limit followed by diagonalization of quasiparticle Hamiltonian}

We substitute Eqs.~(\ref{eqn:PsiBw2})-(\ref{eqn:SplitTrans1}) into the Hamiltonian, Eq.~(\ref{eqn:KBog2}), then take the long-wavelength limit while translating the exponential (crystal) momentum factors to coincide with the Dirac point. The continuum limit effectively converts the sublattice sums into integrals. By performing one of the integrations, over the A sublattice, say, while adhering to nearest neighbor overlaps, we obtain the affective Hamiltonian for the condensate and quasiparticles $\hat{H}= \hat{K}_0 \, +\, \hat{K}_2$ where
\begin{eqnarray}
\fl \hat{K}_0  =  \int \!  d^2r \, \left[  i \hbar c_l \psi^*_A({ \bf r}) \,\mathcal{D}\, \psi_B( { \bf r}) + i \hbar  c_l \psi^*_B( {\bf r })\, \mathcal{D}^* \,\psi_A( { \bf r}) +  \frac{U}{2}  \left|  \psi_A( { \bf r })\right|^4 +  \frac{U}{2} \left|  \psi_B({ \bf r})\right|^4                  \right]  \,   , \\
\fl \hat{K}_2  =  \sum_{j,k}^{\;\;\;\;\;\prime}   \int d^2r\left\{   \hat{\alpha}_j  \hat{\beta}_k^\dagger\,   \hbar  c_l \,  v_{j, A} \, \mathcal{D}^* v^*_{k, B} \, +\,   \hat{\beta}_j  \hat{\alpha}_k^\dagger\, \hbar c_l \, v_{j, B} \, \mathcal{D} v^*_{k, A} \, \right. \nonumber  \\
\fl +  \left.  2 \, U \, \hat{\alpha}_j  \hat{\alpha}_k^\dagger\,   v_{j, A} \left| \psi_A \right|^2  v^*_{k, A}\,+ 2 \, U \, \hat{\beta}_j  \hat{\beta}_k^\dagger\, v_{j, B} \left| \psi_B \right|^2  v^*_{k, B} \right.  \nonumber \\
\fl - \left.   \frac{1}{2} \, U \left| \psi_A \right|^2 \, \hat{\alpha}_j  \hat{\alpha}_k^\dagger\, ( u_{j,A} \, v^*_{k,A} \, +\, u^*_{k,A}  \, v_{j,A} ) -   \frac{1}{2} \, U \left| \psi_B \right|^2 \, \hat{\beta}_j  \hat{\beta}_k^\dagger\, ( u_{j,B} \, v^*_{k,B} \, +\, u^*_{k,B}  \, v_{j,B} )  \right. \nonumber \\
\fl +  \left.  \hat{\alpha}_j^\dagger  \hat{\beta}_k  \,  \hbar   c_l  \,  u^*_{j, A} \, \mathcal{D}^* u_{k, B} \, +\, \hat{\beta}_j^\dagger  \hat{\alpha}_k \, \hbar  c_l \, u^*_{j, B} \, \mathcal{D} u_{k, A} \,  \right. \nonumber\\
\fl+ \left. 2 \, U \,  \hat{\alpha}_j^\dagger  \hat{\alpha}_k \, u^*_{j, A} \left| \psi_A \right|^2  u_{k, A}\, +\, 2 \, U \,  \hat{\beta}_j^\dagger  \hat{\beta}_k \,u^*_{j, B} \left| \psi_B \right|^2  u_{k, B}  \right. \nonumber \\
\fl -\left.  \frac{1}{2} \, U \left| \psi_A \right|^2 \,  \hat{\alpha}_j^\dagger  \hat{\alpha}_k \,( v^*_{j,A} \, u_{k,A} \, +\, v_{k,A}  \, u^*_{j,A} ) -  \frac{1}{2} \, U \left| \psi_B \right|^2 \,  \hat{\beta}_j^\dagger  \hat{\beta}_k \,( v^*_{j,B} \, u_{k,B}  \, +\, v_{k,B}  \, u^*_{j,B} ) 
     \right. \nonumber \\
\fl -\left.    \hat{\alpha}_j \hat{\beta}_k \,  \hbar  c_l  \, v_{j, A} \, \mathcal{D}^* u_{k, B} \, -\, \hat{\beta}_j \hat{\alpha}_k \,  \hbar c_l \, v_{j, B} \, \mathcal{D} u_{k, A} \right.   \nonumber\\
\fl  - \left. \, 2 \, U           \hat{\alpha}_j \hat{\alpha}_k      \, v_{j, A} \left| \psi_A \right|^2  u_{k, A}\, -  2 \, U \, \hat{\beta}_j \hat{\beta}_k\, v_{j, B} \left| \psi_B \right|^2  u_{k, B}    \right. \nonumber \\
\fl + \left.  \frac{1}{2} \, U \left| \psi_A \right|^2 \, \hat{\alpha}_j \hat{\alpha}_k( u_{j,A} \, u_{k,A} \, +\, v_{k,A}  \, v_{j,A} ) +    \frac{1}{2} \, U \left| \psi_B \right|^2     \hat{\beta}_j \hat{\beta}_k\, ( u_{j,B} \, u_{k,B} \, +\, v_{k,B}  \, v_{j,B} )  \right.    \nonumber \\
\fl -  \left. \hat{\alpha}^\dagger_j \hat{\beta}^\dagger_k  \hbar  c_l \, u_{j, A}^* \, \mathcal{D}^* v_{k, B}^* \, -  \,  \hbar  c_l \, \hat{\beta}^\dagger_j \hat{\alpha}^\dagger_k \,u_{j, B}^* \, \mathcal{D} v_{k, A}^* \right. \nonumber\\
\fl -  \left. 2 \, U\, \hat{\alpha}^\dagger_j \hat{\alpha}^\dagger_k  \, u_{j, A}^* \left| \psi_A \right|^2  v_{k, A}^*\, -\, 2 \, U \, \hat{\beta}^\dagger_j \hat{\beta}^\dagger_k \, u_{j, B}^* \left| \psi_B \right|^2  v_{k, B}^* \right. \nonumber \\
\fl + \left.    \frac{1}{2} \, U \left| \psi_A \right|^2\, \hat{\alpha}^\dagger_j \hat{\alpha}^\dagger_k  \, ( u^*_{j,A} \, u^*_{k,A} \, +\, v^*_{k,A}  \, v^*_{j,A} ) + \frac{1}{2} \, U \left| \psi_B \right|^2 \, \hat{\beta}^\dagger_j \hat{\beta}^\dagger_k \, ( u^*_{j,B} \, u^*_{k,B} \, +\, v^*_{k,B}  \, v^*_{j,B} ) \right\} .  \label{eqn:BDGK2}
\end{eqnarray}
Here we have defined the condensate two-spinor in terms of the A and B sublattice components $\Psi({\bf r}) \equiv \left[ \psi_A({\bf r}), \psi_B({\bf r}) \right]^T$, and the Dirac operator is defined as $\mathcal{D} \equiv \partial_x + i \partial_y$. Next, we isolate the first six terms (terms with the daggered operator to the right) and write them as a matrix contraction of two pure operator valued vectors
\begin{eqnarray}
   \left(
  \begin{array}{ c }
        \hat{\alpha}_j , \, \hat{\beta}_j 
       \end{array} \right)     
    \left(   \begin{array}{c c}
                A_{u,v}  &    D_{A,B}  \\
                 D_{B,A}    &  B_{u,v}  
       \end{array} \right)  
   \left( 
  \begin{array}{ c }
        \hat{\alpha}^\dagger_k   \\
         \hat{\beta}^\dagger_k
       \end{array} \right)  \, , \label{eqn:matrix1} \\
  \end{eqnarray}
where
\begin{eqnarray}
A_{u,v}  \equiv  2 U   v_{j, A} \left| \psi_A \right|^2  v^*_{k, A}   -  \frac{1}{2} U \left| \psi_A \right|^2  ( u_{j,A}  v^*_{k,A} + u^*_{k,A}  v_{j,A} ) \, ,\\
B_{u,v}  \equiv  2 U  v_{j, B} \left| \psi_B \right|^2  v^*_{k, B}  - \frac{1}{2}  U \left| \psi_B \right|^2 \, ( u_{j,B}  v^*_{k,B} + u^*_{k,B}  v_{j,B} )\, , \\
D_{A,B}  \equiv  \hbar  c_l   v_{j, A}  \mathcal{D}^* v^*_{k, B}  \, , \\
D_{B,A}  \equiv  \hbar c_l v_{j, B}  \mathcal{D} \,  v^*_{k, A} \,  .
\end{eqnarray}
The eigenvalues are then obtained by
\vspace{0pc}
\begin{eqnarray}
    \mathrm{det} \!  \left(    \begin{array}{c c}
                A_{u,v} \,- \lambda  &   D_{A,B}  \\
                 D_{B,A}  \,    &   B_{u,v}\, -\, \lambda 
       \end{array} \right)  =   0 \,  \\
       \Rightarrow       \nonumber    \\
 ( A_{u,v} \,- \lambda )\,( B_{u,v}\, -\, \lambda )\, -\, D_{A,B} \, D_{B,A}  \, =\, 0 \, ,\\
    \lambda_\pm  =  \frac{\left( A_{u,v} \, +\, B_{u,v} \right)}{2}   \pm \, \frac{1}{2} \sqrt{ \left( A_{u,v} \, -\, B_{u,v} \right) \, +\, 4\, D_{A,B} \, D_{B,A} } \, , 
\end{eqnarray}
and the corresponding eigenvectors follow
\begin{eqnarray}
 \vec{V}_\pm  =  \left( 
  \begin{array}{ c }
         1    \\
         \frac{D_{B,A}}{( \lambda_\pm  - B_{u,v})  }
       \end{array} \right)  \, .
\end{eqnarray}
The unitary matrix that diagonalizes Eq.(\ref{eqn:matrix1}) is
\begin{eqnarray}
 {\bf U}  =  \frac{1}{\sqrt{2}} \left( 
  \begin{array}{ c c }
         1   &        1      \\
         \frac{D_{B,A}}{( \lambda_+  - B_{u,v})  }   &   \frac{D_{B,A}}{( \lambda_-  - B_{u,v})  } 
       \end{array} \right)   \, .
\end{eqnarray}
The first six terms in Eq.~(\ref{eqn:BDGK2}) may be expressed in the new basis as
\begin{eqnarray}
\lambda_{+\{jk\}} \, \hat{c}_{+,j} \,  \hat{c}^\dagger_{+, k} \, +\, \lambda_{- \{jk\}} \, \hat{c}_{-,j} \,  \hat{c}^\dagger_{-, k} \, ,
\end{eqnarray}
where we have included the $j , k$ subscripts on the eigenvalues to be fully descriptive. The new quasiparticle operators can be written in terms of the old ones as
\begin{eqnarray}
 \hat{c}^\dagger_{\pm,j} \, = \frac{1}{\sqrt{2}} \, \left[ \hat{\alpha}^\dagger_j \, +\,   \frac{D^*_{B,A}}{( \lambda^*_{\pm \{jk\}} \, -\, B^*_{u,v})  } \, \hat{\beta}^\dagger_j    \right] \,.
\end{eqnarray}
Note that the right hand side is $k$-dependent which is implied on the left. The substance of the transformation is contained in the momentum and space-dependent eigenvalues
\begin{eqnarray}
\fl \lambda_{\pm\{jk\}}  =  U   v_{j, A} \left| \psi_A \right|^2  v^*_{k, A} -   \frac{1}{4}  U \left| \psi_A \right|^2  ( u_{j,A}  v^*_{k,A}  + u^*_{k,A}   v_{j,A} ) \\
\fl +  U   v_{j, B} \left| \psi_B \right|^2  v^*_{k, B}   -   \frac{1}{4} \, U \left| \psi_B \right|^2  ( u_{j,B}  v^*_{k,B} + u^*_{k,B}  v_{j,B} )  \nonumber \\
\fl \pm \left\{ \left[ U    v_{j, A} \left| \psi_A \right|^2  v^*_{k, A} -  \frac{1}{4} \, U \left| \psi_A \right|^2  ( u_{j,A}  v^*_{k,A} + u^*_{k,A}   v_{j,A} ) \right. \right.  \nonumber \\
\fl  -  \left. \left.  U    v_{j, B} \left| \psi_B \right|^2  v^*_{k, B}  +   \frac{1}{4}  U \left| \psi_B \right|^2  ( u_{j,B}  v^*_{k,B}  + u^*_{k,B}  v_{j,B} )  \right]^2   \right.   \nonumber \\
\fl +   \left.(\hbar c_l)^2 \, v_{j, A} (\mathcal{D}^*  v^*_{k, B})  v_{j, B} (\mathcal{D}  v^*_{k, A})   \right\}^{1/2}\,  \nonumber  .    \label{eqn:Lambda}
\end{eqnarray}

The next step is to constrain the quasiparticle amplitudes in Eq.~(\ref{eqn:Lambda}) (the $u$'s and $v$'s) in order to diagonalize the Hamiltonian with respect to the momentum indices $j$ and $k$.  First, we let
\begin{eqnarray}
\hbar c_l  v_{j, A}\, \mathcal{D}^*  v^*_{k, B}  =  2 U   v_{j, A} \left| \psi_A \right|^2  v^*_{k, A} -   \frac{1}{2} \, U \left| \psi_A \right|^2  ( u_{j,A}  v^*_{k,A}  +  u^*_{k,A}   v_{j,A} ) \nonumber \\
 \hbar c_l  v_{j, B} \, \mathcal{D}\,  v^*_{k, A}   =  2 U   v_{j, B} \left| \psi_B \right|^2  v^*_{k, B} -  \frac{1}{2}  U \left| \psi_B \right|^2  ( u_{j,B}  v^*_{k,B}  + u^*_{k,B}  v_{j,B} )\, ,  \label{eqn:mixedconst}
\end{eqnarray}
and then substitute these into Eq.(\ref{eqn:Lambda}), which reduces the two eigenvalues to
\begin{eqnarray}
\fl \lambda_{+\{jk\}}  = - \, \mu \,  v_{j, A} \,   v^*_{k, A} \, +\, 2 \, U \,   v_{j, A} \left| \psi_A \right|^2  v^*_{k, A} -  \frac{1}{2} \, U \left| \psi_A \right|^2 \, ( u_{j,A} \, v^*_{k,A} \, +\, u^*_{k,A}  \, v_{j,A} ) \nonumber \\
                    \fl  - \mu \,  v_{j, B} \,   v^*_{k, B} \, +\,2 \, U \,   v_{j, B} \left| \psi_B \right|^2  v^*_{k, B}   -   \frac{1}{2} \, U \left| \psi_B \right|^2 \, ( u_{j,B} \, v^*_{k,B} \, +\, u^*_{k,B}  \, v_{j,B} )\, ,   \nonumber  
  \end{eqnarray}                   
and
\begin{eqnarray}
\lambda_{-\{jk\}}  =  0 \, ,
\end{eqnarray}
where we have reinserted the chemical potential terms. It is important that Eq.~(\ref{eqn:mixedconst}) depend only on one index so that quasiparticle amplitudes for different eigeneneregies are not coupled. Dividing Eq.~(\ref{eqn:mixedconst}) through by $v_{j, A} $ and $v_{j, B}$, respectively, cancels all $j$-index terms except for ones that appear as $u_{j,A} / v_{j, A}$ and $ u_{j,B} / v_{j, B}$. To completely decouple the $j$-$k$ modes, we must ensure that $u_{j,A} / v_{j, A} \, =\, u_{j,B} / v_{j, B} = \eta({ \bf r}\,)$, i.e., the amplitudes for any given quasiparticle mode have the same relative spatial form. We can then rewrite $\lambda_{+\{jk\}}$ as
\begin{eqnarray}
 \fl   \lambda_{+\{jk\}} =  \\
 \fl \, \frac{1}{2} \, \hbar  c_l  \, v_{j, A}  \mathcal{D}^*  v^*_{k, B}  - \frac{1}{2} \mu \,  v_{j, A} \,   v^*_{k, A}   +   U   v_{j, A} \left| \psi_A \right|^2  v^*_{k, A} -    \frac{1}{4} \, U \left| \psi_A \right|^2  ( u_{j,A}   v^*_{k,A}  +u^*_{k,A}  v_{j,A} ) \nonumber \\
   \fl   + \, \frac{1}{2} \, \hbar c_l \,  v_{j, B} \, \mathcal{D} \, v^*_{k, A} \,-\, \frac{1}{2}\,  \mu \,  v_{j, B} \,   v^*_{k, B}+ \,  U   v_{j, B} \left| \psi_B \right|^2  v^*_{k, B}   -  \frac{1}{4} \, U \left| \psi_B \right|^2  ( u_{j,B}  v^*_{k,B} + u^*_{k,B}   v_{j,B} )\label{eqn:lambda2}\,.  \nonumber
   \end{eqnarray}
 Finally, we impose the constraints
\begin{eqnarray}
\fl  -\, \frac{1}{4} E_k  v^*_{k, A}  =   \frac{1}{4} \hbar c_l  \mathcal{D}^* v^*_{k,B}  - \frac{1}{4}  \mu  v^*_{k,A} + \frac{1}{2}  U \left| \psi_A \right|^2  v^*_{k,A} - \frac{1}{4} U \left| \psi_A \right|^2 u^*_{k,A} , \label{eqn:Dcond1}  \\
\fl  -  \frac{1}{4}  E_j  v_{j, A}   =    \frac{1}{4}  \hbar  c_l \mathcal{D}^*   v_{j,B}  - \frac{1}{4}  \mu   v_{j,A}  +  \frac{1}{2} U  \left| \psi_A \right|^2  v_{j,A} - \frac{1}{4} U \left| \psi_A \right|^2 \, u_{j,A} \,\label{eqn:Dcond2} . 
\end{eqnarray}
Multiplying Eqs.~(\ref{eqn:Dcond1})-(\ref{eqn:Dcond2}) by $v_{j, A}$ and $v^*_{k,A}$, respectively, and using the property that $\int d^2r  \, v_{j, B} \, \mathcal{D} \, v^*_{k, A} \, =\, \int d^2r ( \mathcal{D}^*v_{j, B}) \,  v^*_{k, A} $ , we may separate out $1/2$ of each derivative term in Eq.~(\ref{eqn:lambda2}), which reduces the non-derivative terms in the first line of Eq.~(\ref{eqn:lambda2}) to
\begin{eqnarray}
-\frac{1}{4} ( E_k \, +\, E_j)\,v^*_{k,A}\, v_{j, A}  \;  .
\end{eqnarray}
We may reduce the second line using the other half of each derivative term, thereby condensing the eigenvalues down to
\begin{eqnarray}
\lambda_{+\{jk\}} \, =\, -\frac{1}{4} ( E_k \, +\, E_j)\,( v^*_{k,A}\, v_{j, A} \, +\, v^*_{k,B}\, v_{j, B} ) \; .\label{eqn:lambda2cond}
\end{eqnarray}
The next six terms in Eq.(\ref{eqn:BDGK2}) may be diagonalized in a similar way yielding the eigenvalues
\begin{eqnarray}
\lambda_{+\{jk\}}  = - \, \mu \,  u^*_{j, A} \,   u_{k, A} \, +\, 2 \, U \,   u^*_{j, A} \left| \psi_A \right|^2  u_{k, A} \nonumber \\ 
  - \frac{1}{2} \, U \left| \psi_A \right|^2 \, ( v^*_{j,A} \, u_{k,A} \, +\, v_{k,A}  \, u^*_{j,A} ) \nonumber \\
   - \mu \,  u^*_{j, B} \,   u_{k, B} \, +\,2 \, U \,   u^*_{j, B} \left| \psi_B \right|^2  u_{k, B}  \nonumber \\
     - \frac{1}{2} \, U \left| \psi_B \right|^2 \, ( v^*_{j,B} \, u_{k,B} \, +\, v_{k,B}  \, u^*_{j,B} )   \, , 
     \end{eqnarray}
   and 
   \begin{eqnarray}
\lambda_{-\{jk\}}  =  0 \, .  \label{eqn:Dcond4} 
\end{eqnarray}
Following our previous steps, we obtain
\begin{eqnarray}
\lambda_{+\{jk\}} \, =\, \frac{1}{4} ( E_k \, +\, E_j)\,( u^*_{k,A}\, u_{j, A} \, +\, u^*_{k,B}\, u_{j, B} ) \; . \label{eqn:lambda3cond}
\end{eqnarray}
Combining Eqs.~(\ref{eqn:lambda2cond}) and~(\ref{eqn:lambda3cond}), and inserting the quasiparticle operators, reduces the first twelve terms in Eq.~(\ref{eqn:BDGK2}) to the expression
\begin{eqnarray}
 \fl \frac{1}{4}  \sum_{j,k}^{\;\;\;\;\;\prime} \!   \int \!  d^2r \, ( E_j +   E_k) \hspace{0pc}  \left[ \hat{c}_{+,j}^\dagger  \hat{c}_{+, k}  ( u^*_{k,A} u_{j, A} + u^*_{k,B} u_{j, B} ) -   \hat{c}_{+,j} \hat{ c}_{+, k}^\dagger ( v^*_{k,A}  v_{j, A} +  v^*_{k,B} v_{j, B} ) \right]. 
\end{eqnarray}
 For the special case where $j\, =\, k$, we may further combine the terms at the cost of an extra c-number term to arrive at
\begin{eqnarray}
  \fl  - \frac{1}{2}  \sum_{k}^{\;\;\;\;\;\prime}   \int \! d^2r\, 2  E_k ( v^*_{k,A}  v_{k, A}  + v^*_{k,B} v_{k, B} ) \nonumber \\
  \fl +  \frac{1}{4}  \sum_{j,k}^{\;\;\;\;\;\prime}   \int \! d^2r  ( E_j + E_k) \,  \hat{c}_{+,j}^\dagger  \hat{c}_{+, k} ( u^*_{k,A} u_{j, A} - v^*_{k,A}  v_{j, A}   + u^*_{k,B} u_{j, B} - v^*_{k,B}  v_{j, B} ) .  \label{eqn:87} 
\end{eqnarray}
Applying the completeness relations $ \int \! d^2r\, ( u^*_{k,A} u_{j, A} -  v^*_{k,A}v_{j, A} )$$=$$\delta_{i, j}$ and $\int \! d^2r\, ( u^*_{k,B} u_{j, B} - v^*_{k,A} v_{j, B} )$$=$$\delta_{i, j}$, contracts Eq.~(\ref{eqn:87}) down to
\begin{eqnarray}
\fl -  \sum_{k}^{\;\;\;\;\;\prime}   \int d^2r\,  E_k \,( |v_{k,A}|^2 + | v_{k, B}|^2 )  +    \sum_{k}^{\;\;\;\;\;\prime}   E_k \, \hat{c}_{+,k}^\dagger  \hat{c}_{+, k}  \, . 
\end{eqnarray}
Diagonalizing the rest of Eq.~(\ref{eqn:BDGK2}) (terms with no daggered operators and ones with only daggered operators) by capitalizing on the $j$-$k$ symmetry of terms such as $\int \!  d^2r \,  u_{k, A} \, v_{j, A}$, and anti-symmetry of the $(E_j \, -\, E_k)$ factor, we obtain the final form of the interacting Hamiltonian
\begin{eqnarray}
\fl \hat{H}   =   \int \! d^2r  \left[  i \hbar c_l\, \psi^*_A({ \bf r }) ( \partial_x  +  i \partial_y ) \psi_B({ \bf r}) \, +\,   i \hbar c_l \, \psi^*_B({ \bf r} ) ( \partial_x  -  i \partial_y ) \psi_A({ \bf r})  \right. \\
+\left.   U/2  \left|  \psi_A( { \bf r})\right|^4 +  U/2\left|  \psi_B({ \bf r})\right|^4                  \right]  \nonumber \\
        - \sum_{j}^{\;\;\;\;\;\prime}   E_j   \!     \int \! d^2r   \,{\bf v}^{T*}_j  {\bf v}_j   +  \sum_{j}^{\;\;\;\;\;\prime} \,  E_j      \,         \hat{c}^\dagger_{+ , \, j}  \, \hat{c}_{+, \, j} \, ,  \label{eqn:HRLSE}
\end{eqnarray}
with the resulting constraints on quasiparticle amplitudes given by
\begin{eqnarray}
\hbar  c_l \mathcal{{D}^*} u_{j,B} + \left( 2  U \left| \psi_A \right|^2 - \mu \right) u_{j,A} - U \left| \psi_A \right|^2 v_{j,A}   =  E_j u_{j, A} \, ,   \label{eqn:RLSE1}\\
\hbar  c_l \mathcal{{D}} \, u_{j,A}  + \left(  2 U \left| \psi_B \right|^2 - \mu \right) u_{j,B} - U\, \left| \psi_B \right|^2  v_{j,B}  =  E_j  u_{j, B} \, ,  \label{eqn:RLSE2}    \\
- \hbar c_l  \mathcal{D} \, v_{j,B}  - \left(  2U \left| \psi_A \right|^2 - \mu \right)  v_{j,A} +   U \left| \psi_A \right|^2 \, u_{j,A}  =  E_j  v_{j, A} \, ,     \label{eqn:RLSE3} \\
- \hbar  c_l  \mathcal{D^*}  v_{j,A} -  \left( 2U \left| \psi_B \right|^2 - \mu \right) v_{j,B}  +  U\, \left| \psi_B \right|^2 u_{j,B}  =   E_j  v_{j, B}\,  .  \label{eqn:RLSE4} 
\end{eqnarray}

\subsection{Second method: Diagonalize quasiparticle Hamiltonian then impose tight-binding}

Although the first method is cumbersome it is the more rigorous approach and instils confidence in the final constraint equations. A shorter approach is to first obtain the usual Bogoliubov equations for a condensate not confined in a lattice, and then apply the tight-binding limit directly. The Bogoliubov Hamiltonian is
\begin{eqnarray}
\fl \hat{H} =  \int \!  d^2r \,  \Psi^*({ \bf r }) \left[ H_0 - \mu  + \frac{g}{2}   \left| \Psi({ \bf r }) \right|^2\right] \Psi({ \bf r })  -\sum_j^{\; \;\;\;\;\prime}  E_j  \! \int\!  d^2r \left| v_j({ \bf r}) \right|^2  +  \sum_j^{\; \;\;\;\;\prime} E_j \alpha_j^\dagger \alpha_j  \, , \label{eqn:finalform}                           
\end{eqnarray}
with the constraint equations (BdGE) given by
\begin{eqnarray}
\mathcal{L} \, u_j \, -\, g \,  \left| \Psi \right|^2 \, v_j  =  E_j u_j \label{eqn:BDG1}  \\
\mathcal{L}^* \, v_j \, -\, g \,  \left| \Psi \right|^2 \, u_j  =  -\, E_j v_j  \label{eqn:BDG2}\, . 
\end{eqnarray}
In Eqs.~(\ref{eqn:BDG1})-(\ref{eqn:BDG2}), $\mathcal{L}$ is a differential operator containing terms that couple the quasiparticle and condensate velocities. An additional implicit constraint is that $\Psi$ satisfies the nonlinear Schr\"odinger equation. To pass to the tight-binding limit we express all spatial functions in Eqs.~(\ref{eqn:finalform})-(\ref{eqn:BDG2}) in terms of Wannier functions for the individual sublattices, and evaluate the Bloch plane wave factors at the Dirac point momemtum. Adhering to nearest-neighbor overlap for on-site Wannier functions, we integrate out spatial degrees of freedom (which splits the honeycomb lattice into A and B sublattices), regroup terms into finite differences, and then take the continuum limit. Equation~(\ref{eqn:finalform}) then transforms to Eq.~(\ref{eqn:HRLSE}), while Eqs.~(\ref{eqn:BDG1})-(\ref{eqn:BDG2}) transform to Eqs.~(\ref{eqn:RLSE1})-(\ref{eqn:RLSE4}) with several additional derivative terms contained in $\mathcal{L}$ as follows
\begin{eqnarray}
-\frac{\hbar^2}{2m} \left[   \nabla^2 +  i \nabla^2 \phi  + 2 i \,  \nabla \phi \cdot \nabla - ( \nabla \phi )^2 \right] u_j \, ,  \label{eqn:BDGkinetic}
\end{eqnarray}
where $\phi$ is the condensate phase. After going through the steps that culminate in the tight-binding continuum limit, these terms transform to
\begin{eqnarray}
\fl i\hbar  c_l \mathcal{D}^*u_{k, B(A)}  + \left[ - i \hbar \tau_1   \nabla  \phi_{A(B)} \cdot  \nabla +  \hbar \tau_2 | \nabla  \phi_{A(B)} | - i \hbar \tau_3 ( \nabla^2 \phi_{A(B)}) \right]  u_{k,A(B)}  \, ,
\end{eqnarray}
where the coefficients encapsulate the spatial integrals as follows:  $\tau_1 \!\! \propto | \int \! d{\bf r}\,  w_A^* \nabla w_B |$,  $\tau_2 \! \! \propto | \int \! d{\bf r} \, w_A^* (\nabla \phi) w_A |$, $\tau_3\!\! \propto \int \! d{\bf r} \, w_A^* |\nabla \phi|^2 w_A $. These extra terms depend on the condensate phase $\phi_{A(B)}$, and so couple the superfluid velocity to the quasiparticle excitations. In particular, the term with coefficient $\tau_1$ depends on the direction of quasiparticle emission relative to the motion of the condensate. The relativistic linear stability equations, Eqs.~(\ref{eqn:RLSE1})-(\ref{eqn:RLSE4}), may be expressed in compact notation as
\begin{eqnarray}
\left( \begin{array} {cc} 
    \tilde{{D}}        &      - \Delta  \\
           \Delta    &      -  \tilde{{D}}^*
           \end{array} \right)      \left( \begin{array} {c} 
                       {\bf u}_{\bf k}        \\
                        {\bf v}_{\bf k} 
           \end{array} \right)   =     \tilde{ E}_{\bf k}   \left( \begin{array} {c} 
                       {\bf u}_{\bf k}        \\
                        {\bf v}_{\bf k} 
           \end{array} \right)  \,  ,  \label{finalcompact}
\end{eqnarray} 
where
\begin{eqnarray}
   \Delta  \equiv  U  \, \mathrm{diag}(    \left| \psi_A \right|^2 ,   \left| \psi_B \right|^2  ) \, ,  \\   
   \tilde{ E}_{\bf k}  \equiv  E_{\bf k} \cdot   \mathbb{1}_2  \, ,    \\
            {[}  \tilde{{D} } {]}_{1,1}   \equiv  - \mu + 2 U \left| \psi_A \right|^2 - i \hbar \tau_1  \nabla \phi_A \cdot \nabla  +  \hbar \tau_2 \left|\nabla \phi_A \right| - i \hbar \tau_3 \left( \nabla^2 \phi_A\right) \, ,  \\
                    {[}  \tilde{{D} } {]}_{2,2}   \equiv  - \mu + 2 U \left| \psi_B \right|^2 - i\hbar \tau_1   \nabla \phi_B \cdot \nabla  +  \hbar \tau_2 \left| \nabla \phi_B \right| - i \hbar \tau_3 \left( \nabla^2 \phi_B\right) \, ,  \\
 {[} \tilde{{D} }  {]}_{1,2}  =  {[} \tilde{{D} }  {]}_{2,1}^*\,  \equiv \,  i \hbar c_l  \mathcal{D}^* \,, \label{finalcompact2}
\end{eqnarray}
and $  \mathbb{1}_2$ is the $2 \times 2$ unit matrix.

\section{Stability of vortex solutions}
\label{Stability}

\begin{figure}[h]
\centering
\subfigure{
\label{fig:ex3-a}
 \includegraphics[width= 1\textwidth ]{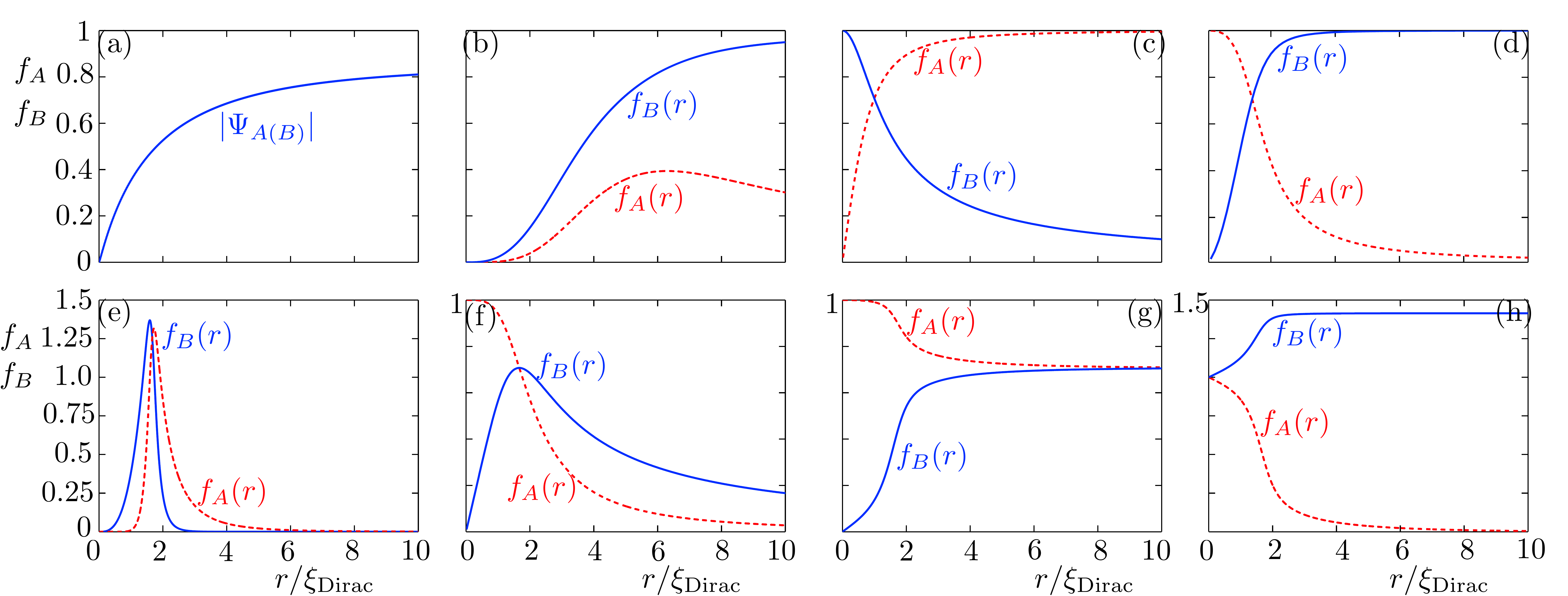}}   \\
\caption[]{(color online) \emph{NLDE vortex radial solutions}. (a) Bessel solution for $\ell=3$; (b) numerical solution for $\ell =4$; (c) vortex/soliton; (d) Anderson-Toulouse vortex; (e) ring-vortex solution for $\ell=4$; (f) ring-vortex/soliton solution; (g) Mermin-Ho vortex; (h) half-quantum vortex. In each plot, the upper and lower spinor radial solutions are indicated in red and blue, respectively. }
\label{Radial2}
\end{figure}

Two independent derivations of the RLSE in Sec.~\ref{RLSE} and their reduction to the BdGE, which we discuss in Sec.~\ref{Reductions}, establishes Eqs.~(\ref{finalcompact})-(\ref{finalcompact2}) as the correct method for computing the low-energy structure (quasiparticle states and eigenenergies) for arbitrary vortex solutions of the NLDE~\cite{Haddad2012}. Radial profiles for vortex solutions of the NLDE are plotted in Fig.~\ref{Radial2}, with details of our solution methods presented in~\cite{Haddad2015}. Briefly, the plots in Fig.~\ref{Radial2} were obtained by solving the NLDE in plane-polar coordinates for the case of cylindrical symmetry, which read
\begin{eqnarray}
- \hbar c_l \! \left(\! \partial_{r } +  \frac{\ell}{r} \right)\! f_B(r) + U \left|f_A(r)\right|^2\! f_A(r)    = \mu  f_A(r) \label{eqn:CondPsi7} \, ,  \\
  \hbar c_l\! \left( \!\partial_{r }+ \frac{1\! - \! \ell }{r} \right)\! f_A(r) + U  \left|f_B(r)\right|^2 \! f_B(r)  =   \mu f_B(r)  \label{eqn:CondPsi8}\,  . 
\end{eqnarray}
for the upper and lower two-spinor component radial functions $f_A$ and $f_B$, and $\ell \in \mathbb{Z}$ the angular quantum number coming from imposing single-valuedness of the wavefunction. 
The most immediate and pragmatic concern is the combined effect of the honeycomb lattice geometry and the inter-particle interaction on the lifetime of a vortex. It should be emphasized that the presence of an infinite tower of negative energy states below the Dirac point seems to imply that a condensate residing there will eventually decay provided there is a mechanism for energy dissipation into noncondensate modes (i.e., secondary interactions with thermal atoms).$^{\footnotemark[2]}$ \footnotetext[2]{We remind the reader that this infinite tower of negative energy states is only in the Dirac cone approximation.} Generically, negative energy states are present for moving condensates for which excitations subtended by a backward cone have negative frequencies~\cite{Fetter1972}. Moreover, when a vortex is present small displacements of the core from the symmetry axis of the trap results in a precession of the core, which, when combined with dissipation, causes the vortex to spiral to the edge of the condensate. In the absence of a periodic lattice potential this dynamical process is known to be driven by the anomalous modes in the linear spectrum, i.e., modes with negative energy and positive norm~\cite{feder2000} also called Goldstone modes. The time for a vortex to spiral to the edge of the trap would then define its lifetime. In the absence of the lattice this precessional motion is canceled by introducing rotation to the trap~\cite{feder2000,Fetter2001}, a result which we suspect to be true in the lattice case as well.

To undertake a full treatment of the lifetime would mean computing this spiraling time and then comparing it with the lifetime that we compute here due to the dynamical instability from the complex frequencies. The lifetime of the vortex would then be the smaller of the two values. Nevertheless, in cases where dissipation is weak and the vortex is centered on the symmetry axis of the trap, the dominant source of instability arises from the complex eigenvalues associated with RLSE modes. We will limit our analysis to the effect of the latter, and regard the negative real part of the eigenvalues from a standpoint of metastability. Physically, the complex eigenvalue gives rise to fluctuations in the angular rotation of the vortex spinor components~\cite{haddad2011}. In the case of the NLDE this is a result of internal ``friction'', i.e., energy exchange, between the two spinor components displayed in the complex derivative terms of the Dirac kinetic energy. This drag force between the two vortex components (or between vortex and soliton) eventually causes substantial degradation of the vortex. This is the measure that we will use to compute vortex lifetimes.


\subsection{Numerical solution of the relativistic linear stability equations and vortex lifetimes}

\begin{figure}[h]
\centering
\subfigure{
\label{fig:ex3-a}
\hspace{-.3in} \includegraphics[scale=.4]{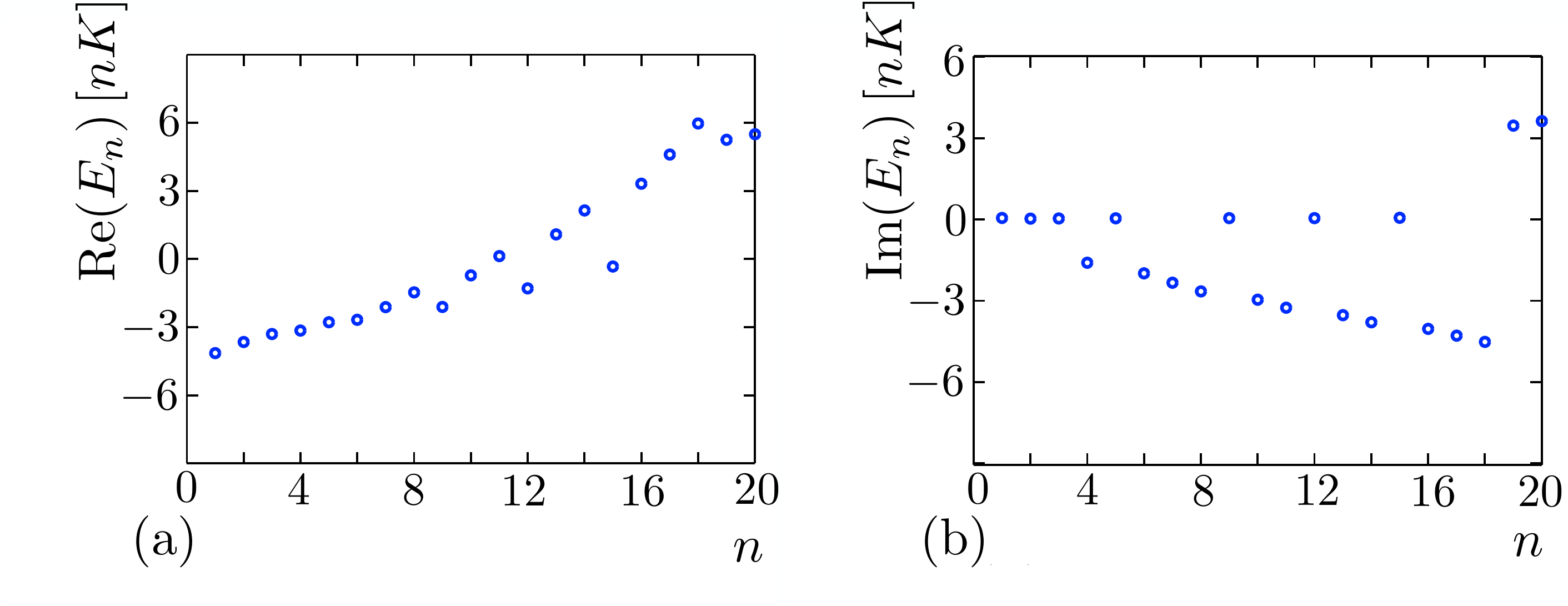}} \\
\caption[]{(color online) \emph{Anomalous mode frequencies for the vortex/soliton}. The real part of the anomalous mode frequencies are plotted in (a), the Imaginary parts are plotted in (b). The horizontal axis labels the excitation mode determined by the quasiparticle angular momentum quantum number $n$.}
\label{Anom}
\end{figure}

The stability of a particular condensate density and phase profile such as a vortex is arrived at by expanding Eq.~(\ref{finalcompact}) and expressing differential operators in terms of suitable coordinates, for example polar coordinates for a vortex, then using separation of variables for the quasiparticle amplitudes with the appropriate form of $\psi_{A(B)}$, i.e., solutions of Eqs.~(\ref{eqn:CondPsi7})-(\ref{eqn:CondPsi8}). This yields a set of first-order coupled ODE's in the radial coordinate to be solved consistently for the functions $u_{A(B)}(r)$, $v_{A(B)}(r)$ and the eigenvalues $E_k$. We discretize the derivatives and functions using a forward-backward average finite-difference scheme, then solve the resulting discrete matrix eigenvalue problem using MATLAB function Eig. In Fig.~\ref{Anom} we have plotted the real and imaginary parts of the first 20 eigenvalues, labeled by the quantized quasiparticle rotation number $n \in \mathbb{Z}$, for the vortex/soliton solution which we discuss in previous work~\cite{Haddad2012}. The lowest modes are anomalous with negative real parts and positive, nonzero but small, imaginary parts. 

\begin{figure}[h]
\centering
\subfigure{
\label{fig:ex3-a}
\hspace{-.2in} \includegraphics[scale=.4]{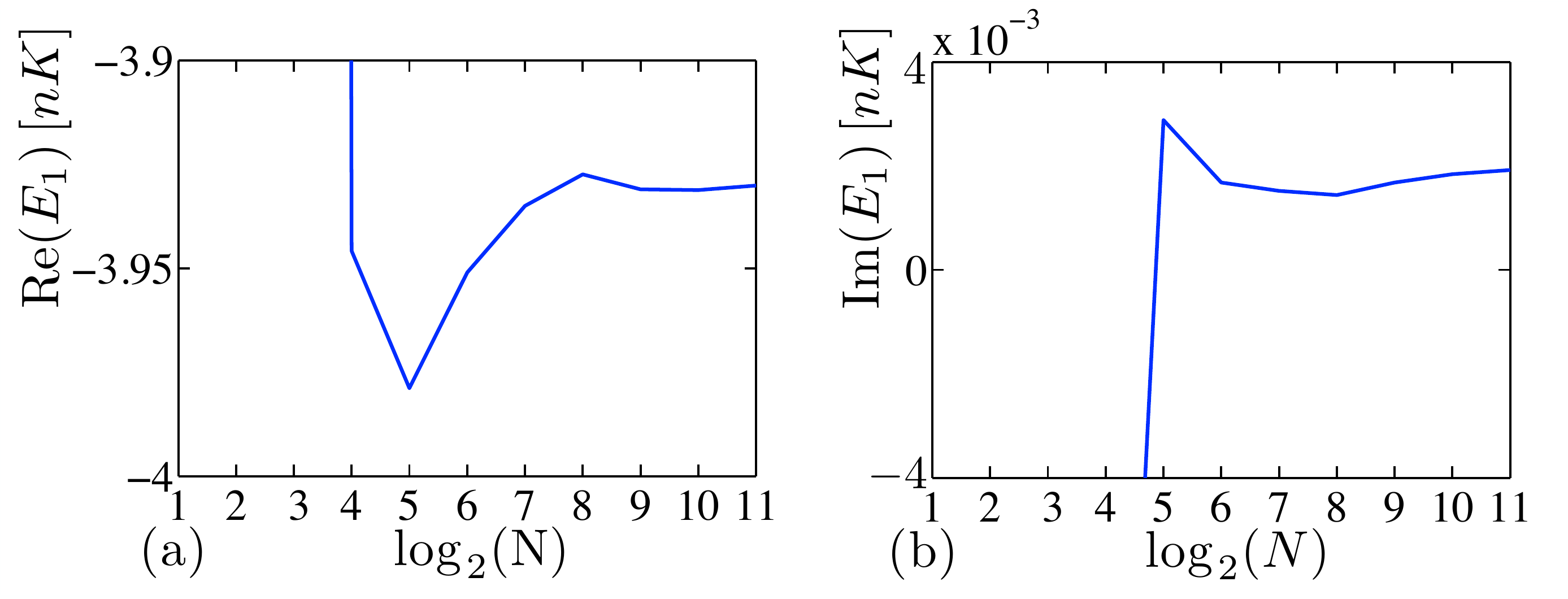}}   \\
\caption[]{(color online) \emph{Convergence of RLSE for the vortex/soliton}. Real (a) and imaginary (b) parts of the lowest anomalous mode. The horizontal axis shows the number of steps and the energy of the lowest excitation of the vortex/soliton corresponding to $n=1$ in Fig.~\ref{Anom} is plotted on the vertical axis. }
\label{AnomConv}
\end{figure}

Convergence of RLSE eigenvalues for the $l=1$ vortex/soliton background as a function of the grid size $N$ used in the $4N \times 4N$ matrix problem is displayed in Fig.~\ref{AnomConv}, where we have plotted the real and imaginary parts of the eigenvalue for the lowest excitation mode. The lifetime of a particular vortex solution can be computed by examining the lowest quasiparticle rotation mode $n =1$, since at very low temperatures this mode dominates the spectrum. The lifetime is then characterized by the reciprocal of the imaginary part of the associated eigenvalue, i.e., $\mathrm{lifetime} \equiv \hbar/\mathrm{Im}\left(E_{-1}\right)$. Here, the $-1$ subscript refers to quasiparticle rotation relative to the vortex rotation. Eigenvalues for the lowest quasiparticle rotational mode and the associated lifetimes for all of our solutions are listed in Table~\ref{vortexstability}. 

\begin{table}[h]
\begin{center}
\resizebox{14.5cm}{!}{ \begin{tabular}{ c ||  c     c       }
\hline  \vspace*{-3mm}  \\
Solution type  & \, Quasiparticle energy $[ \mathrm{nK}] \;\; (\pm 10^{-4} \mathrm{nK})$ &\, \hspace{1pc}  Lifetime $[\mathrm{s}]$   \\
\hline \hline \vspace{-3mm}  \\
Complex topological vortex   & $ 2.231 -  4.174 \times 10^{2} i$ & \hspace{2pc} $1.969 \times 10^{-5} \;  (\pm 4 \times 10^{-12})$       \\
\hline  \vspace*{-3mm}  \\
Topological vortex   &  $ 8.184 \times 10^{-3} - 1.066 \times 10^{3} i $   & \hspace{2pc} $1.941 \times 10^{-5} \;  (\pm 6 \times 10^{-13})$       \\
\hline  \vspace*{-3mm}  \\
Ring-vortex  & $- 4.181 -  1.599 \times 10^{-2} i $  & $0.5295 \;  (\pm 3 \times 10^{-3})$        \\
\hline  \vspace*{-3mm}  \\
\, Ring-vortex/soliton  \,  &$- 4.203 +  2.022 \times 10^{-3} i $ & \, $4.043 \;  (\pm 2 \times 10^{-1})$         \\
\hline  \vspace*{-3mm}  \\
Vortex/soliton  &  \hspace{0pc}$-4.211  +  2.141 \times 10^{-3}  i$      &\,  $3.841 \;  (\pm 2 \times 10^{-1})$            \\
\hline \vspace*{-3mm}  \\
Mermin-Ho vortex  & \hspace{0pc}  $2.818 \times 10^2 + 1.066 \times 10^5 i$ & \,\hspace{2.25pc}  $7.712 \times 10^{-8} \;  (\pm 7 \times 10^{-17})$             \\
\hline \vspace*{-3mm}  \\
Anderson-Toulouse vortex  &  \hspace{0pc} $- 4.202 + 2.033 \times 10^{-3} i$ & \, $4.041  \;  (\pm 2 \times 10^{-1})$          \\
\hline \vspace*{-3mm}  \\
Half-quantum vortex  &  \hspace{0pc}$2.818 \times 10^2 + 1.066 \times 10^5 i$ & \, \hspace{2pc} $7.712 \times 10^{-8}  \;  (\pm 7 \times 10^{-17})$           \\ \hline
\end{tabular}}
{\caption{\emph{Stability properties of NLDE vortices.} Lifetimes are computed using the value of the interaction $U$ in Table~\ref{table1} and the formula $\mathrm{lifetime} \, = \,  \hbar/\mathrm{Im}\left(E_{-1}\right)$. The various vortex solutions of the NLDE are plotted in Fig.~\ref{Radial2} and derived in detail in~\cite{Haddad2015}. Note that solutions with similar boundary conditions have lifetimes of the same order of magnitude.      }   \label{vortexstability}}
 \end{center}
\end{table}

\begin{figure}[h]
\centering
\subfigure{
\label{fig:ex3-a}
\hspace{-.2in} \includegraphics[scale=.25]{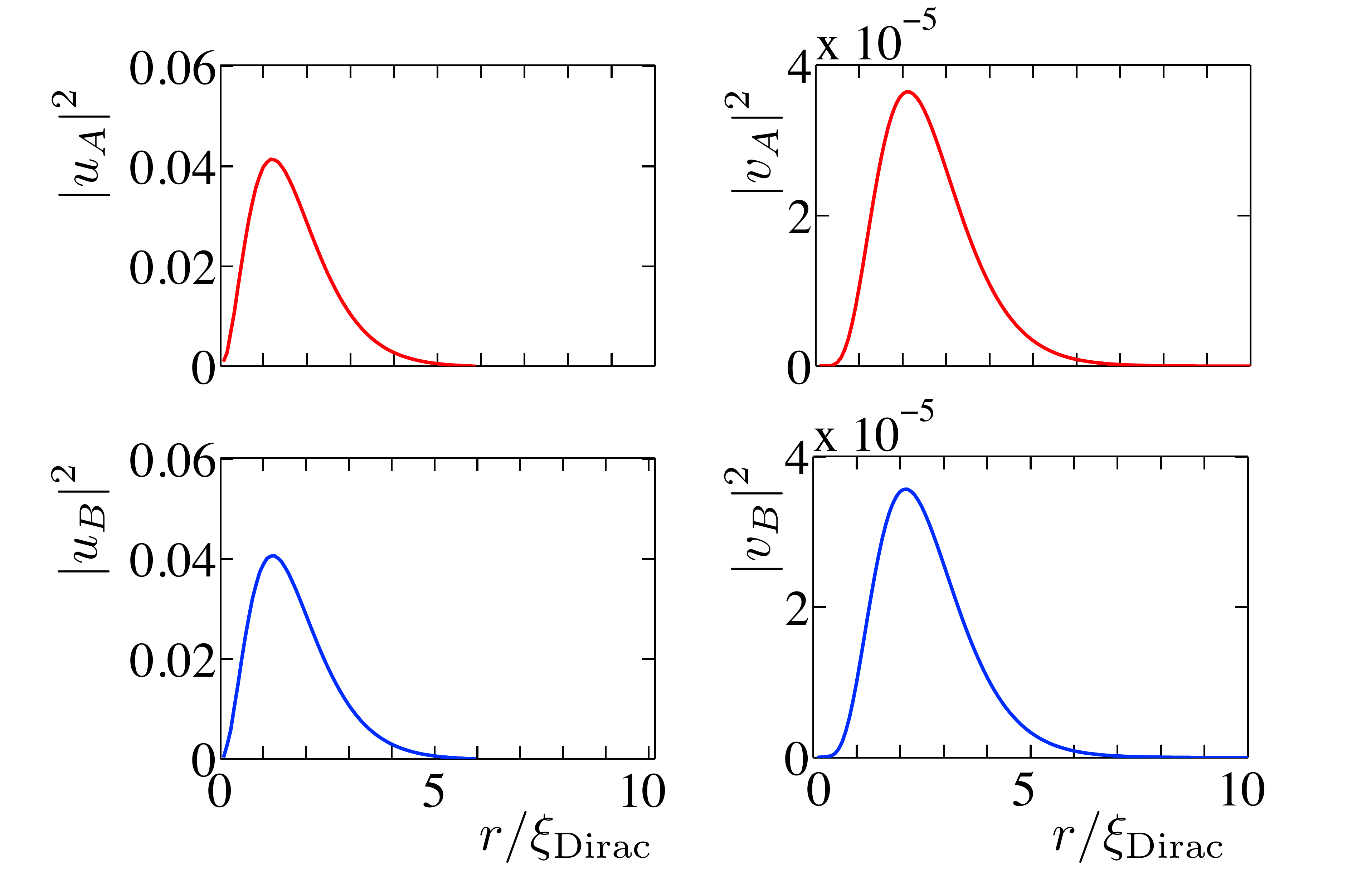}}   \\
\caption[]{(color online) \emph{Plots of lowest quasiparticle excitation for the vortex/soliton configuration.} Used with permission~\cite{haddad2011}. }
 \label{VortexSolitonExcitation}
\end{figure}

To understand the character of the quasiparticle modes we must consider the spatial functions associated with each eigenvalue. Radial profiles for the $n=1$ quasiparticle excitation in the vortex/soliton background are shown in Fig.~\ref{VortexSolitonExcitation}. They are bound states near the core of the vortex localized specifically at the point where the soliton and vortex components of the background are equal (see Fig.~\ref{Radial2}(c)). 
Physically, the imaginary part of the eigenvalues imply a transfer of energy between the vortex and soliton components through quantum fluctuations. In particular, each component acquires quantum admixtures from different rotational modes as well as local shifts in amplitude from phase and density fluctuations, respectively. Mathematically, the full quasiparticle operator with time and spatial dependence for this mode is $\hat{\phi}({\bf r}, t) \simeq \left[ \hat{\phi}_{A,-1}({\bf r}, t), \, \hat{\phi}_{B,-1}({\bf r }, t) \right]^T$, where the quasiparticle spinor operators are 
\begin{eqnarray}
\hspace{-2pc}   \hat{\phi}_{A,-1}({\bf r}, t) =    e^{-iE_{-1} t/\hbar} \, e^{-i \theta } \, u_{A, -1 }(r)       \, \hat{\alpha}_{-1} + e^{ iE_{-1} t/\hbar} \, e^{-i \theta } \,  v^*_{A,-1}(r)             \, \hat{\alpha}^\dagger_{-1}  \, ,   \\
 \hspace{-2pc}   \hat{\phi}_{B,-1}({\bf r}, t) =  e^{-iE_{-1} t/\hbar}\,  u_{B,-1}(r)    \,                    \hat{\beta}_{-1} + e^{iE_{-1} t/\hbar} \, v^*_{B,-1}(r)     \,                 \hat{\beta}^\dagger_{-1}  \,  .
\end{eqnarray}
As discussed previously, relative to the vortex background the quasiparticle has rotation $\ell = - n = -1$, which has the effect of reducing the rotation of the vortex. Note that the expression for the operator $\hat{\phi}({\bf r}, t)$ is approximate since we have truncated the sum over quasiparticle modes after the lowest mode. We recall that the spatial functions have the properties $u_{A, -1 }(r), \, u_{B, -1 }(r) \sim 10^{-2}$ and $v_{A, -1 }(r), \, v_{B, -1 }(r) \sim 10^{-5}$~(see Ref.~\cite{haddad2011}), where all are peaked in the ``notch'' region $\xi_{\mathrm{Dirac}} < r < 2\xi_{\mathrm{Dirac}}$, and where the absolute values of the slopes of the soliton and vortex are maximum. In this region, the normalization integrals (one for each sublattice) are given by
\begin{eqnarray}
 \int \! d^2r \, \left[  |u_{A(B), -1 }(r)|^2 - |v_{A(B), -1 }(r)  |^2  \right]  > 0 \, . 
 \end{eqnarray}
 This combination of positive norm and negative $\mathrm{Re}(\mathrm{E}_{-1})$ signals the presence of the anomalous mode. In Sec.~\ref{Reductions}, we will see that these bound quasiparticle modes solve the Majorana equation, which predicts an additional zero energy mode localized at the same distance from the center of the vortex.

%


\section{Connection to other theories}
\label{Reductions}

In this section we examine several reductions of the RLSE to other equations familiar to BECs, superconductivity, graphene, and high energy physics. Our results demonstrate the variety of substructures  contained within the RLSE framework. Note that we adhere to the weakly interacting regime through all of our derivations as explained in Sec.~\ref{Constraints}.


\subsection{Reductions of the relativistic linear stability equations}
\label{ReducedTheories}

Mappings of the RLSE to other more fundamental equations proceeds by resolving the RLSE solution space into a parameterization with respect to two measures: 1) the ratio of the on-site energy, which we denote $t_0$, to the chemical potential $\mu$; and 2) the ratio of the quasiparticle kinetic energy, $c_l p$, to the boson interaction strength $U$. The on-site energy is calculated as the average value of the Laplacian plus lattice potential over same-site Wannier functions. As such, $t_0$ encodes the sublattice energy offset into a uniform mass gap at the Dirac points of the continuum theory. Changing $t_0/\mu$ tunes the spectrum between a linear (relativistic) and quadratic (nonrelativistic) dispersion. In contrast, changing $c_l p/U$ from large to small values tunes the spectrum from a pure particle-like dispersion to one characterized by an equal admixture of particles and holes. The dimensionality of solutions to Eq.~(\ref{finalcompact}) experiences a corresponding change over the parameter space from a single-component Schr\"odinger-like solution (for $t_0/\mu , \, c_lp/U \sim 1$) to a four-component spinor solution (for $t_0/\mu , \,  c_lp/U  \ll 1$), the latter similar to Nambu-Gorkov states in superconductors arising from doubling of  Fermion degrees of freedom. Thus, the quasiparticle spectrum is spanned by a two-dimensional parameter space highlighting the similarity between Bogoliubov and relativistic structures.

To quantify our discussion we look for a non-relativistic reduction of Eq.~(\ref{finalcompact}) by working first from the lattice form of the NLDE since the hopping terms are the same as those for the RLSE. We recall that the standard massive Dirac equation has a well defined non-relativistic limit to the Schr\"odinger equation~\cite{bjorken64}. The proof uses the fact that in the low-energy limit the mass term, proportional to $m c^2$, is the largest contribution to the energy. In particular, the two-spinor formulation of the Dirac equation is comprised of two coupled equations. The procedure involves isolating the mass term in one equation and  substituting the resulting expression into the second equation. The substitution converts the first-order spatial gradient to a second-order Schr\"odinger kinetic term for small relative kinetic to mass energy. This effectively pushes the mass dependence out into smaller correction terms which may be neglected. Similar steps may be implemented in our case but we must first introduce an offset between the sublattice potential well depths (a mass gap) so that we obtain the desired curvature in the spectrum near the Dirac points, effectively introducing a non-relativistic regime.

\begin{figure}[h]
     \begin{center}
     \subfigure{
          \includegraphics[width=.95\textwidth]{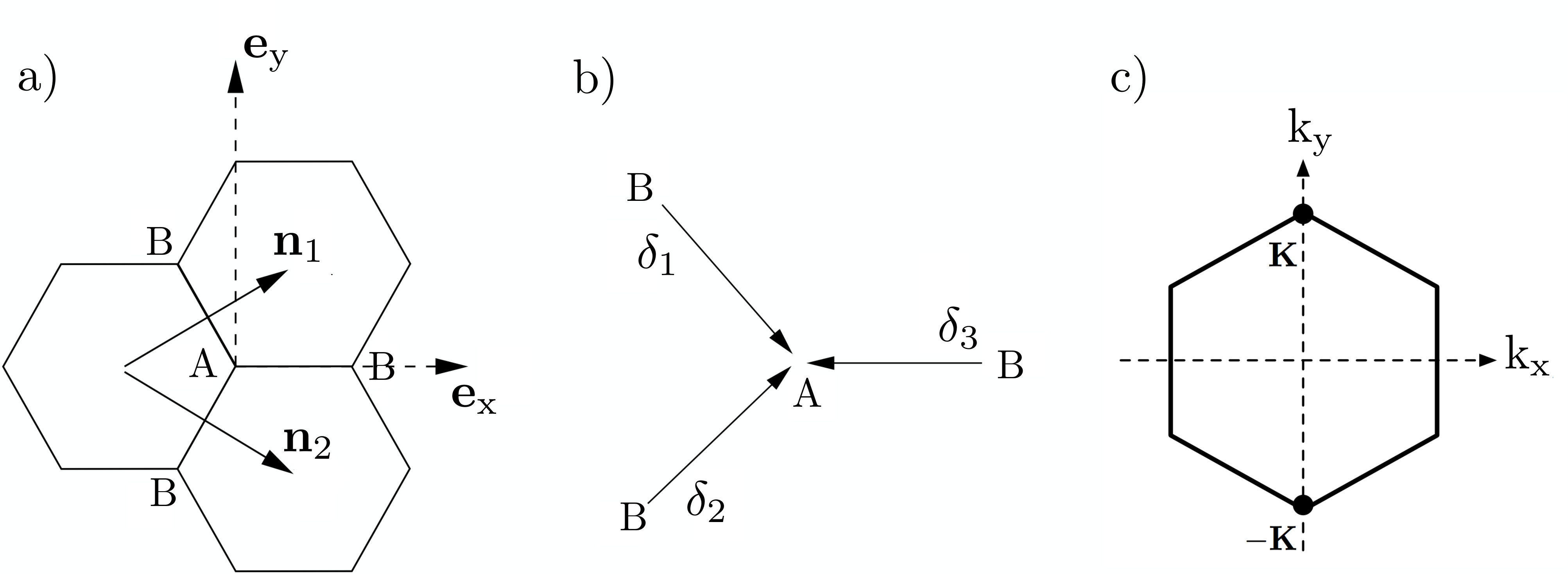}}
          \end{center}
          \caption[]{(color online) \emph{Characterization of a honeycomb lattice}. a) Hexagonal lattice structure. b) Nearest neighbor displacement vectors. c) Reciprocal lattice. Used with permission~\cite{haddad2009}.   }
\label{Honeycomb_Lattice}
\end{figure}

Starting from the discrete NLDE for a single Dirac point and following similar steps as in our previous work~\cite{haddad2009}, we obtain 
\begin{eqnarray}
\fl \mu   \psi_{A_{ j}}   =  - t_h \left( \psi_{B_{  j}}  e^{i {\bf k }\cdot { \bf \delta}_3 }+  \psi_{B_{  j-n_1}}  e^{i { \bf k}\cdot { \bf \delta}_1 } + \psi_{B_{ j-n_2}} \,   e^{i {\bf k}\cdot {\bf \delta}_2 }\right) -  t_0   \psi_{A_{j} }  + U  \left|  \psi_{A_{ j}} \right|^2  \,  \psi_{A_{ j}}  \, , \label{discretenlde1} \\
\fl  \mu   \psi_{B_{ j}}       =  - t_h \left( \psi_{A_{  j}} e^{- i { \bf k}\cdot { \bf \delta}_3 } \,+\, \psi_{A_{  j+n_1}}  e^{- i { \bf k}\cdot {\bf \delta}_1 }+ \psi_{A_{  j+n_2}}    e^{- i { \bf k}\cdot  {\bf \delta}_2 }\right)  +  t_0    \psi_{B_{ j} }  + U  \left|  \psi_{B_{ j}} \right|^2   \psi_{B_{  j} }  \, , \label{discretenlde2}
\end{eqnarray}
where $t_h$,  $t_0$, $U$, and ${\bf k}$ are the hopping, same-site, and on-site interaction energies and crystal momentum, respectively. The ${\bf \delta}$'s, ${\bf n}$'s, and 2D vector indices ${\bf j}$ indicate the lattice vectors described in Fig.~\ref{Honeycomb_Lattice}. In Eqs.~(\ref{discretenlde1})-(\ref{discretenlde2}), $t_0$ is the sublattice offset equivalent to a spectral gap $2 \vert  t_0 \vert$. For weak interactions, the on-site energy can be made much larger than the contact interaction strength by tuning the lattice potential so that $| \mu \pm  t_0 |>> U$. After inserting the correct values for the lattice vectors and solving Eq.~(\ref{discretenlde2}) for $\psi_{B_j}$, to zeroth-order in $U/\vert \mu -  t_0 \vert$ we obtain
\begin{eqnarray}
 \psi_{B_{ \bf j}}  = -   \frac{ t_h}{ \mu -  t_0   }  \left( \psi_{A_{ j}}  + \psi_{A_{  j+n_1}}   e^{ i 2\pi/3 }  + \psi_{A_{ j+n_2}}   e^{- i 2\pi/3}\right)   \label{DNLDB2}  \, .
\end{eqnarray}
From Eq.~(\ref{DNLDB2}) we may write analogous expressions for neighboring sites by shifting the indices using the lattice vectors ${\bf n}_j$,
\begin{eqnarray}
   \psi_{B_{ j-n_1}}  =  -  \frac{ t_h}{  \mu -  t_0   }  \left(   \psi_{A_{  j-n_1}}+  \psi_{A_{ j} } e^{ i 2\pi/3 } + \psi_{A_{ j+ (n_2 - n_1)}}  e^{- i 2\pi/3}\right)  \, ,  \label{DNLDB3}  \\
  \psi_{B_{  j-n_2}}  =  -   \frac{ t_h}{  \mu - t_0  }  \left(  \psi_{A_{  j-n_2}} + \psi_{A_{  j-(n_2 - n_1)}} e^{ i 2\pi/3 } + \psi_{A_{ j}}    e^{- i 2\pi/3}\right)  \label{DNLDB4}  .
\end{eqnarray}
Substituting Eqs.~(\ref{DNLDB2})-(\ref{DNLDB4}) into Eq.~(\ref{discretenlde1}), expanding complex factors and regrouping the terms to form finite differences, we arrive at the expression
\begin{eqnarray}
 \fl  \mu    \psi_{ j}     =   \frac{ t_h^2}{ 2 ( \mu -  t_0 )}  \left\{  \left( \psi_{  j+n_1} - 2  \psi_{j}  +  \psi_{ j-n_1}  \right) + \left( \psi_{ j+n_2}  - 2   \psi_{ j}  +  \psi_{j-n_2}  \right) \right. \nonumber  \\
 \fl + \left.  \right(  \psi_{ j+ (n_2 - n_1)}  - 2  \psi_{ j} +  \psi_{ j- (n_2 - n_1)}      \left)    - i \sqrt{3} \left[   \left( \psi_{j+n_1}   -  \psi_j  \right) +  \left( \psi_{ j}  -  \psi_{ j-n_1} \right)  -  \left( \psi_{ j+n_2}  - \psi_{ j} \right)      \right. \right. \nonumber \\
 \fl -   \left. \left. \left( \psi_{ j} - \psi_{ j-n_2} \right)  +   \left( \psi_{ j+ (n_2 - n_1)}  -  \psi_{ j} \right)   +  \left( \psi_{ j}   - \psi_{ j-(n_2 - n_1)} \right) \right] \right\}  -   t_0   \psi_{ j}   + U  \left|  \psi_{ j} \right|^2    \psi_{ j}     \, . \label{eqn:DNLS}  
\end{eqnarray}
Equation~(\ref{eqn:DNLS}) is a discrete nonlinear Schr\"odinger equation for the honeycomb lattice in the sense that it has as its continuum limit the usual nonlinear Schr\"odinger equation with cubic nonlinearity. Substituting the correct continuum forms for the finite differences and then expressing the result in rectangular coordinates, we obtain 
\begin{eqnarray}
\fl   \mu  \psi   =  \\
\fl -  \frac{ t_h^2   a^2}{ 2 ( \mu -  t_0 ) }  \left[   \left( \frac{3}{4}  \frac{\partial^2}{\partial x^2} +  \frac{1}{4}  \frac{\partial^2}{\partial y^2} - \frac{\sqrt{3}}{2} \frac{\partial^2}{ \partial x \partial y } \right)  \psi  + \left( \frac{3}{4}  \frac{\partial^2}{\partial x^2} + \frac{1}{4}  \frac{\partial^2}{\partial y^2} +  \frac{\sqrt{3}}{2}   \frac{\partial^2}{\partial x \partial y }\right) \psi \right.   \nonumber  \\
\fl  +  \left.  \frac{\partial^2\psi}{\partial y^2} -  i \frac{\sqrt{3}}{a} \left(  \sqrt{3} \frac{\partial\psi}{\partial x} -  \frac{\partial\psi}{\partial y} - \sqrt{3} \frac{\partial\psi}{\partial x} - \frac{\partial\psi}{\partial y}  + 2\,  \frac{\partial \psi}{\partial y}  \right)  \right]   -  t_0   \psi   + U  \left|  \psi \right|^2    \psi    \label{eqn:NLS}   \, , \nonumber 
\end{eqnarray}
which finally reduces to
\begin{eqnarray}
\left(  \mu +  t_0 \right)  \psi   =  -  \frac{ c_l^2 \hbar^2 }{  ( \mu -  t_0  )} \nabla^2 \psi    + U  \left|  \psi \right|^2   \psi       \, ,  \label{NLDEtoNLSE1}
\end{eqnarray}
where we have substituted in the effective speed of light $c_l = \sqrt{3} t_h a/2\hbar$. Performing the same steps with Eq.~(\ref{discretenlde2}) gives a second equation 
\begin{eqnarray}
 \left(  \mu - t_0 \right)  \psi   =  - \frac{ c_l^2 \hbar^2 }{  ( \mu  +  t_0  )} \nabla^2 \psi    + U  \left|  \psi \right|^2   \psi       \, .   \label{NLDEtoNLSE2}
\end{eqnarray}

Next we examine two limits. For $t_0 \ll \mu$, Eqs.~(\ref{NLDEtoNLSE1})-(\ref{NLDEtoNLSE2}) describe two propagating modes where the effective mass and total energy are of the same order. To lowest order in $t_0 / \mu$ Eqs.~(\ref{NLDEtoNLSE1})-(\ref{NLDEtoNLSE2}) become 
\begin{eqnarray}
 \mu^2   \psi  +   c_l^2 \hbar^2  \,  \nabla^2 \psi   + \mu t_0 \psi  - \mu U  \left|  \psi \right|^2   \psi   = 0   \, , 
 \end{eqnarray}
and
 \begin{eqnarray}
  \mu^2   \psi  +   c_l^2 \hbar^2  \,  \nabla^2 \psi   -   \mu t_0 \psi  - \mu U  \left|  \psi \right|^2   \psi   = 0 \, .   \label{NLDEtoNLSE3}
\end{eqnarray}
Reintroducing time dependence through $\mu^2 \to - \hbar^2 \partial^2_t$ and dividing through by $c_l^2 \hbar^2$ gives 
\begin{eqnarray}
 \frac{1}{c_l^2} \frac{\partial^2 \psi }{\partial t^2}   -  \nabla^2 \psi  -   \frac{m^2 c_l^2}{\hbar^2} \psi  + \frac{ \tilde{U}^2 }{c_l^2 \hbar^2}  \left|  \psi \right|^2   \psi    =  0   \, ,    \label{NLDEtoNLSE4} \\
  \frac{1}{c_l^2} \frac{\partial^2 \psi }{\partial t^2}   -  \nabla^2 \psi  +    \frac{m^2 c_l^2}{\hbar^2} \psi  + \frac{\tilde{U}^2 }{c_l^2 \hbar^2}  \left|  \psi \right|^2   \psi    =  0 \, ,  \label{NLDEtoNLSE5}
\end{eqnarray}
where we have defined the mass $m \equiv \sqrt{ \vert \mu t_0 \vert }/c_l^2$ and interaction strength $\tilde{U} \equiv \sqrt{ \vert \mu U \vert }$. Equations~(\ref{NLDEtoNLSE4})-(\ref{NLDEtoNLSE5}) are nonlinear Klein-Gordon equations describing a tachyon mode with imaginary mass in Eq.~(\ref{NLDEtoNLSE4}), and an ordinary Klein-Gordon mode with real mass in Eq.~(\ref{NLDEtoNLSE5}). In contrast, if we tune the lattice potential offset so that $t_0 \sim \mu$ the mode described by Eq.~(\ref{NLDEtoNLSE1}) has a very small effective mass and large energy, whereas the mode in Eq.~(\ref{NLDEtoNLSE2}) will have a very large mass and small energy. In this case the mode in Eq.~(\ref{NLDEtoNLSE2}) gets ``frozen out'' and we are left with only one propagating mode in Eq.~(\ref{NLDEtoNLSE1}). Here multiplication by the total energy $\mu + t_0$ does not cancel the effective mass $\mu - t_0$ in the denominator of the gradient term. Reintroducing the time dependence by $\mu + t_0 \to i \hbar \partial_t$ and the effective mass $m = (\mu - t_0)/2 c_l^2$, Eq.~(\ref{NLDEtoNLSE1}) reduces to the nonlinear Schr\"odinger equation$^{\footnotemark[3]}$ \footnotetext[3]{This step can be justified formally from the Heisenberg equation of motion for the wavefunction starting from the operator formalism, but such justification is well known from theory of NLSE.} 
\begin{eqnarray}
  i \hbar  \frac{\partial \psi }{\partial t}   + \frac{\hbar^2}{2 m}  \nabla^2 \psi   - U  \left|  \psi \right|^2   \psi    =  0   \, .  \label{NLDEtoNLSE6} 
\end{eqnarray}
Thus, tuning $t_0$ interpolates between a Dirac and a Schr\"odinger structure with Klein-Gordon bridging the two. One may understand the intermediate Klein-Gordon result through a general argument by noting that any reduction of the Dirac equation to the Schr\"odinger equation must modify both the single-particle dispersion as well as the spin attached to each excitation mode. To be precise, two regimes are identified: one associated with binding two spin-1/2 modes into a single spin-0 mode at lower energy resolutions (Eqs.~(\ref{discretenlde1})-(\ref{discretenlde2}) to Eqs.~(\ref{NLDEtoNLSE4})-(\ref{NLDEtoNLSE5})), and one associated with a crossover from relativistic to classical dispersion (Eqs.~(\ref{NLDEtoNLSE4})-(\ref{NLDEtoNLSE5}) to Eq.~(\ref{NLDEtoNLSE6})). This observation applies not only to the fundamental case, but also in the honeycomb lattice picture with regards to pseudospin.

Applying the same steps to the RLSE, Eq.~(\ref{finalcompact}) yields the BdGE 
\begin{eqnarray}
     i \hbar \frac{ \partial }{\partial t}  \left( \begin{array} {c} 
                       {  u}      \\
                        { v}
           \end{array} \right)  =   \left( \begin{array} {cc} 
        -  \frac{\hbar^2}{2m} \nabla^2 + \Delta_m     &    \;  - \Delta_p    \\
           \Delta_p    &   \;          \frac{\hbar^2}{2m} \nabla^2 -  \Delta_m  
           \end{array} \right)      \left( \begin{array} {c} 
                        u        \\
                         v 
           \end{array} \right)     \; ,  \label{bogoliubov}  
\end{eqnarray} 
with $\Delta_m =  - \mu +  2 U |\psi|^2$,   $\Delta_p  =   U |\psi|^2$, where $\psi$ is the condensate wavefunction for either of the decoupled sublattices and the effective mass is the same as in Eq.~(\ref{NLDEtoNLSE6}), $m =  (\mu - t_0)/2 c_l^2$ . Note that we have suppressed explicit space-time dependence in Eq.~(\ref{bogoliubov}) for clarity. In the particle regime for large characteristic momentum, $c_l \vert {\bf p}\vert \gg  U$, the particle and hole amplitudes satisfy $u \gg v$ and Eq.~(\ref{bogoliubov}) reduces to the standard Schr\"odinger equation for a particle moving in the potential $V \equiv - \Delta_m$.

Next, we look at the case of single-mode approximation for the pseudospin degrees of freedom in Eq.~(\ref{finalcompact}), i.e., where the sublattice backgrounds are equal $\psi_A \equiv \psi_B$ which also  implies that $u_A= u_B\equiv u({\bf r}, t)$ and $v_A= v_B \equiv v({\bf r}, t)$. One then finds that the system Eq.~(\ref{finalcompact}) reduces to the Andreev equation
\begin{eqnarray}
     i \hbar \frac{ \partial }{\partial t}  \left( \begin{array} {c} 
                       {  u}      \\
                        { v}
           \end{array} \right)  =   \left( \begin{array} {cc} 
        - i  \hbar c_l  \hat{{ \bf p}} \cdot \nabla + \Delta_m     &    \;  - \Delta_p  \\
           \Delta_p    &   \;      i  \hbar c_l  \hat{{\bf p}} \cdot \nabla - \Delta_m 
           \end{array} \right)      \left( \begin{array} {c} 
                        u        \\
                         v 
           \end{array} \right)     \,  .   \label{andreev}         
\end{eqnarray} 
The unit vector $\hat{{\bf p}}$ in Eq.~(\ref{andreev}) points in the direction of quasiparticle propagation. Here we have chosen the case of zero background flow $\nabla \phi_{A, B} = 0$ as we will do for the remainder of this section except for the vortex background. Equation~(\ref{andreev}) is the Andreev equation for propagation through a medium comprised of both normal and superconducting regions~\cite{Andreev1964}. The spatially dependent pairing and mass terms are $\Delta_p({\bf r}) = U |\psi({\bf r})|^2$ and  $\Delta_m({\bf r}) = 2 U |\psi({\bf r})|^2 - \mu$. In this analogy the condensate wavefunction $\psi({\bf r})$ stands in for the order parameter in a superconducting medium. Equation~(\ref{andreev}) describes slowly varying particle and hole functions $u({\bf r})$ and $v({\bf r})$ split off from an overall rapidly oscillating plane wave portion which moves in the direction $\hat{ {\bf p}}$. Thus, we should expect similar exotic scattering such as specular and retro-reflection~\cite{Beenakker2006}.

Next, we look at the particle regime where the particle component is dominant, $u_{A(B)} \gg v_{A, (B)}$. In this regime Eq.~(\ref{finalcompact}) reduces to the Dirac equation 
\begin{eqnarray}
     i \hbar \frac{ \partial }{\partial t}  \left( \begin{array} {c} 
                        u_A      \\
                         u_B 
           \end{array} \right)  =   \left( \begin{array} {cc} 
                           \Delta_A     &    \;  - i \hbar c_l (\partial_x - i \partial_y)     \\
           - i \hbar c_l (\partial_x + i \partial_y)    &   \;           \Delta_B  
           \end{array} \right)      \left( \begin{array} {c} 
                        u_A        \\
                         u_B 
           \end{array} \right)     \; ,  \label{Dirac}     
\end{eqnarray} 
where a potential term appears $\Delta_{A(B)}({\bf r})   =    2 U |\psi_{A(B)}({\bf r})|^2 - \mu$. Equation~(\ref{Dirac}) further reduces to the massless Dirac equation in the case of a constant background $|\psi_{A(B)}|^2 \equiv  \mu/(2U)$.

\begin{figure}[]
\begin{center}
\hspace{0pc} \subfigure{
\label{fig:ex3-a}
\hspace{0in} \includegraphics[scale=1.3]{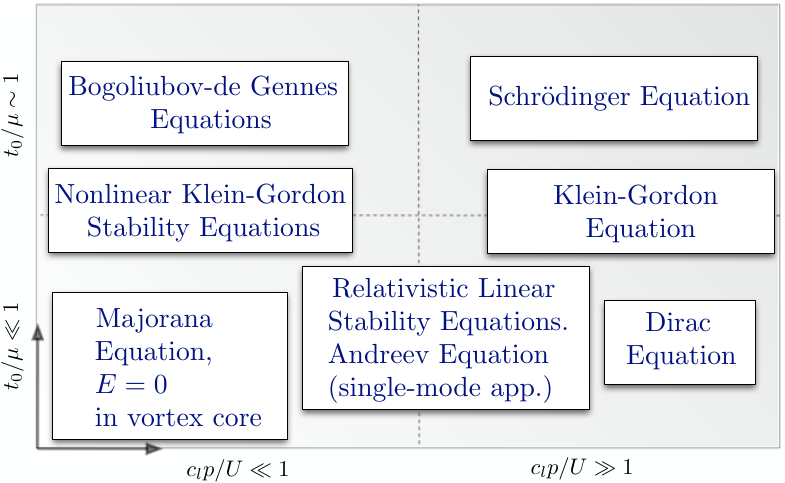}}
\vspace{-1pc}
\end{center}
\caption[]{(color online) \emph{Schematic of reductions of the RLSE}. Limiting theories for the RLSE are displayed for relative strengths of the lattice potential offset and chemical potential along the vertical direction and quasiparticle momentum versus interaction in the horizontal direction. The dashed lines indicate crossovers in the underlying lattice theory.}
\label{RLSE_Regimes}
\end{figure}

Interestingly, zero-mode solutions ($E =0$) of the RLSE occur as well and we find that these solve the Majorana equation which is implicit in the RLSE for certain background configurations. To see this we set $E_k=0$ in Eq.~(\ref{finalcompact}), which decouples the system into two sets of equations in the extreme long-wavelength regime characterized by $|u_{A(B)}| = |v_{A(B)}|$. In this regime Eq.~(\ref{finalcompact}) gives two copies of the form
\begin{eqnarray}
     \left( \begin{array} {cc} 
                           \Delta_A     &    \;  - i \hbar c_l (\partial_x - i \partial_y)     \\
           - i \hbar c_l (\partial_x + i \partial_y)    &   \;           \Delta_B  
           \end{array} \right)      \left( \begin{array} {c} 
                        u_A        \\
                         u_B 
           \end{array} \right)  = 0   \; ,  \label{Majorana1}    \hspace{2pc} \emph{} 
\end{eqnarray} 
where the potential terms are $\Delta_{A(B)}  =    U |\psi_{A(B)}|^2 - \mu$. For a uniform condensate, i.e., far from any vortex cores, the asymptotic choices are $U |\psi_{A(B)}|^2  \to \mu, 0$. In both cases Eq.~(\ref{Majorana1}) offers no solution. However, for the vortex/soliton there is a ``notch'' in the order parameter near the core, where $|\psi_A|^2 =  |\psi_B|^2  <   \mu/U $ $\Rightarrow      \Delta_{A(B)}  < 0$, in which case Eq.~(\ref{Majorana1}) reduces to the Majorana equation
\begin{eqnarray}
- i {\bf \sigma} \cdot \nabla \psi_c  + m  \psi = 0 \, ,   \label{Majorana2}       
\end{eqnarray}
where $\psi_c  =  i \psi^* \equiv  [ u_A^* , \, u_B^* ]$, $m \equiv |\Delta_{A(B)}|$, and $\psi : \mathbb{R}^2 \to \mathbb{R}^2$. Equation~(\ref{Majorana2}) supports real solutions with linear dispersion and has been studied extensively in its original mathematical form~\cite{Majorana1937} and more recently in condensed matter physics intimately associated with topological insulators~\cite{Kane2008}. In their present incarnation these Majorana zero modes also occur in the core of nonlinear Dirac vortices with higher winding ($\ell > 1$ in Ref.~\cite{Haddad2012}) where both spinor components vanish $|\psi_{A(B)}(0, \theta)|^2 = 0$. In this case the mass term in Eq.~(\ref{Majorana2}) reduces to the condensate chemical potential $m = \mu$. In the superfluid context the meaning of the Majorana zero mode is of a zero-energy pure spatial density fluctuation associated with rigid translations of the vortex core. Here phase fluctuations only appear as finite-energy fluctuations in the vortex rotational and translational motion. For the vortex/soliton the zero mode is a circular ring reflecting the symmetry under both rigid rotations as well as translations of the vortex. In Fig.~\ref{RLSE_Regimes} we summarize the various types of reductions of the RLSE indicating the conditions or limits for each equation type.

\subsection{Mapping to relativistic Bardeen-Cooper-Schrieffer theory}
\label{RelativisticBCS}

In this section we discuss the modifications needed to connect the RLSE to relativistic BCS theory. Here we capitalize on an important property of the NLDE and RLSE. This is that repulsive interactions for bosons in the honeycomb lattice break the valley particle-hole exchange symmetry at the Dirac point in a significant way such that an additional sign change of the interaction restores the symmetry. More properly stated, the noninteracting theory is invariant independently under charge conjugation ($\mathcal{C}$), parity inversion ($\mathcal{P}$), and time reversal ($\mathcal{T}$). Repulsive interactions break $\mathcal{T}$ and $\mathcal{C}$, but the symmetry-breaking cancels in such a way as to preserve the full $\mathcal{CPT}$ symmetry~\cite{haddad2009}. Consequently, a parity inverted positive energy solution (valley particle) can be interpreted as a negative energy solution (valley hole) in a theory with attractive interactions but without parity inversion. Stated differently, a theory of particles with repulsive interactions is equivalent to a theory of holes with attractive interactions. 

To complete the mapping to BCS theory we introduce a mass term and nearest-neighbor interactions at the lattice scale to couple the different spinor components. The mass term is obtained through an asymmetry in the honeycomb sublattice potential depths, an intermediate step in populating Dirac points, as we have explained in~\cite{Haddad2012}. The various types of relativistically invariant interactions may be constructed using nearest-neighbor interactions as follows. Specifically, the symmetry of the nonlinearity in the NLDE determines the symmetry of the superconducting order parameter and pair potential in the corresponding BCS analog equations~\cite{Capelle1999-1}. The vector-vector interaction can be obtained by including repulsive nearest-neighbor interactions. A scalar-scalar type coupling can be realized similarly, but by using attractive (instead of repulsive) nearest-neighbor interactions in addition to the repulsive on-site interactions. The spin and pseudo-spin symmetric terms are characterized by an alternating sign for the coupling between the two spinor components. This type of coupling may be realized in a lattice setting via Feshbach resonances using a beam with the proper spatial modulation to produce interactions whose sign alternates between neighboring lattice sites. Pseudo-scalar forms can be realized by eliminating on-site interactions while retaining repulsive nearest-neighbor interactions.

\begin{figure}[h]
\begin{center}
\hspace{0pc} \subfigure{
\label{fig:ex3-a}
\hspace{0in} \includegraphics[scale=1.5]{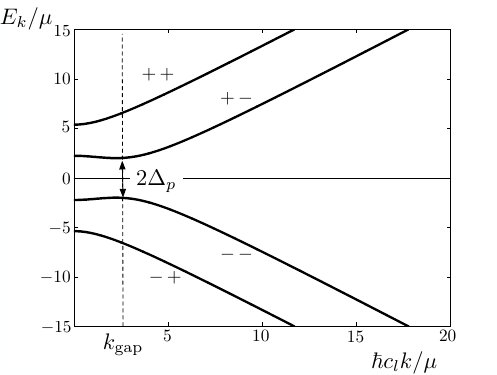}}
\vspace{-1pc}
\end{center}
\caption[]{(color online) \emph{Relativistic BCS spectrum}. Four branches corresponding to the sign combinations in Eq.~(\ref{BCSspectrum}): $++$, $+ -$, $--$, $-+$ (top to bottom). The vertical scale is in units of the chemical potential and the horizontal scale is in units of the reciprocal of the condensate healing length. We have indicated the superconducting gap $2 \Delta_p$ located at $k_\mathrm{gap} $$=$$ (\mu/\hbar c_l) $$\left[ 1+   (2 m_s c_l^2/\mu)^2 \right]^{1/2}$. }
\label{BCS_Plot}
\end{figure}

The case of scalar-scalar coupling in the NLDE with equal on-site and nearest-neighbor interactions $U = U_\mathrm{nn}$ and mass term $m_s c_l^2$ (see ref.~\cite{Haddad2012}) elevate the RLSE to the form
 \begin{eqnarray}
\fl   \tilde{ E}_{\bf k}   \left( \begin{array} {c} 
                       {\bf u}_{\bf k}        \\
                        {\bf v}_{\bf k} 
           \end{array} \right)  =   \label{DBdGE}  \\
           \fl  \left( \begin{array} {cc} 
    i \hbar c_l \sigma \cdot \nabla  + m_s c_l^2 \cdot \mathbb{1}_2  + q  \sigma_\mu  A^\mu &    - i  \Delta_p \sigma_y  \\
            i \Delta_p   \sigma_y   &      -      i \hbar c_l \sigma \cdot \nabla  + m_s c_l^2 \cdot \mathbb{1}_2  + q  \sigma_\mu  A^\mu  
           \end{array} \right)      \left( \begin{array} {c} 
                       {\bf u}_{\bf k}        \\
                        {\bf v}_{\bf k} 
           \end{array} \right)   \nonumber 
\end{eqnarray} 
where $\Delta_p({\bf r}) \equiv U \!  \left[  \vert \psi_A({\bf r}) \vert^2 + \vert \psi_B({\bf r}) \vert^2 \right]$ is the scalar pairing function, the effective polarized 4-vector potential in (2+1) dimensions (so reduced to 3 components) is $A^\mu({\bf r}) \equiv  (U/q)  \left[ \vert \psi_A({\bf r}) \vert^2 - \vert \psi_B({\bf r})   \vert^2 - \mu/U  , \, \vert \psi_A({\bf r}) \vert^2 -  \vert \psi_B({\bf r}) \vert^2 , \, 0 \right]$, and $q$ is an effective charge. As before $\mathbb{1}_2$ is the two-dimensional unit matrix. Equations~(\ref{DBdGE}) comprise the relativistic Bogoliubov-de Gennes equations also known as the Dirac-Bogoliubov-de Gennes equations~\cite{Capelle1999-1,Capelle1999-2}. In the special case of a uniform condensate which solves the nonlinear Dirac equation we have $\vert \psi_A \vert^2 = \vert \psi_B \vert^2  = \mu/U$ and Eq.~(\ref{DBdGE}) yields the eigenvalues
\begin{eqnarray}
E_k = \pm  \sqrt{ \left[  \sqrt{ ( \hbar c_l k)^2 + ( m_s c_l^2)^2} \pm ( m_s c_l^2 + \mu )  \right]^2  + 4 \mu^2    } \,,   \label{BCSspectrum}
\end{eqnarray}
where the magnitude of the quasiparticle momentum $k = \vert {\bf k} \vert$ labels the eigenstates. The signs outside of the radical relate to pseudospin valley states and those inside the radical to the particle-hole Nambu states. The spectrum Eq.~(\ref{BCSspectrum}) is plotted in Fig.~\ref{BCS_Plot}. 
\begin{figure}[b]
\begin{center}
\hspace{0pc} \subfigure{
\label{fig:ex3-a}
\hspace{0in} \includegraphics[width= \textwidth]{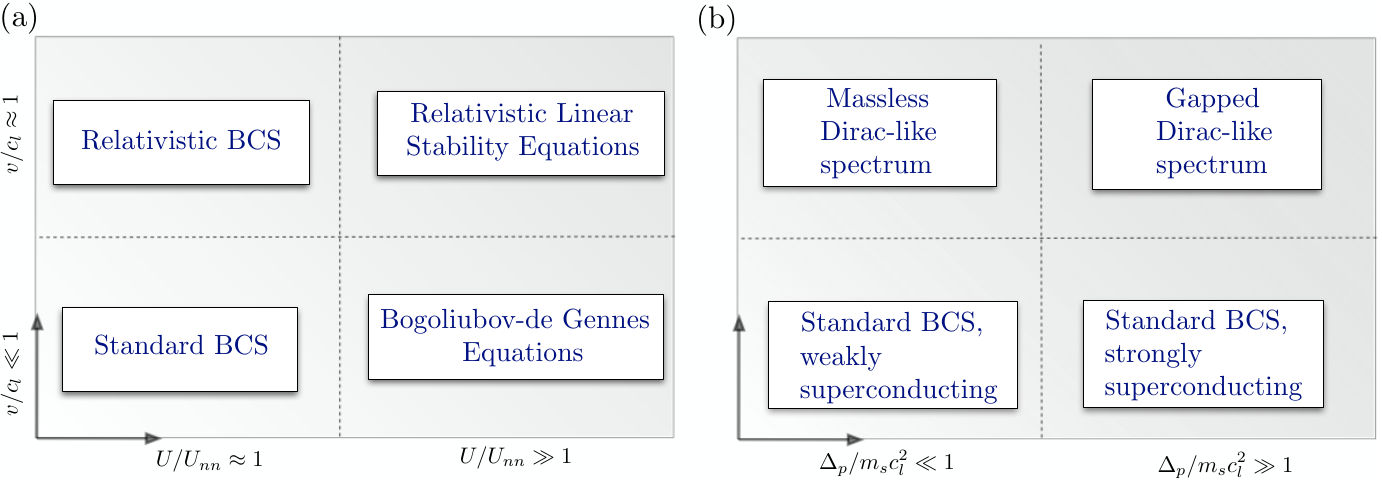}}
\vspace{-1pc}
\end{center}
\caption[]{(color online) \emph{Schematic of regimes for the augmented RLSE}. (a) Limiting theories for relativistic and non-relativistic regimes (vertical axis) versus the relative strengths of on-site and nearest-neighbor interactions (horizontal axis). The size of the superconducting gap is the same as that of the electron-positron spectral gap. (b) Limiting theories for relativistic and non-relativistic regimes (vertical direction) versus the relative strengths of superconducting and electron-positron gaps (horizontal direction). Here the strengths of the on-site and nearest-neighbor interactions are equal. In both (a) and (b) the dashed lines indicate separation between the different regimes and correspond to zero-temperature crossovers in the underlying lattice theory.   }
\label{BCS_Regimes}
\end{figure}

In our BCS analogy the electron-positron spectral gap is $2 m_s c_l^2$ and the superconducting gap is $2 \Delta_p = 4 \mu$ located at $k_\mathrm{gap}^2 $$=$$ k_F^2$$\left[ 1+  4 (c_l/v_F)^2 \right] $. Here the analogs of the Fermi wavenumber and velocity are $k_F \equiv   \mu/\hbar c_l$ and $v_F = \hbar k_F/m_s =   \mu/ m_sc_l$. In our analogy we see that the Fermi momentum is inversely related to the relativistic healing length $\xi = \hbar c_l /\mu$ (defined in terms of chemical potential $\mu$) such that $p_F = \hbar/\xi$. Consider the non-superconducting limit of the spectrum in which $\Delta_p \to  0 \Rightarrow \mu/U  \ll 1$, keeping in mind that in the analog BCS system the Cooper pair mass is $\Delta_p/c_l^2$. Equation~(\ref{BCSspectrum}) reduces to 
\begin{eqnarray}
E_k = \pm \left[   \sqrt{ ( \hbar c_l k)^2 + ( m_s c_l^2)^2} \pm   m_s c_l^2  \right]  \, .  \label{BCSspectrum2}
\end{eqnarray}
Furthermore, for excitations much smaller than the mass gap, $\hbar c_l k \ll m_s c_l^2$, the four branches of the spectrum reduce to two free positive and negative energy Schr\"odinger-like excitations
\begin{eqnarray}
E_k = \pm  \left(\frac{\hbar c_l  }{ 2 m_s c_l } \right)^2   k^2   \, ,   \label{BCSspectrum3}
\end{eqnarray}
and two similar excitations but shifted by constant potentials $\pm 2 m_s c_l^2$
\begin{eqnarray}
E_k = \pm \left[  \left(\frac{\hbar c_l  }{ 2 m_s c_l } \right)^2   k^2  + 2 m_s c_l^2  \right]  \, .    \label{BCSspectrum4}
\end{eqnarray}
Conversely, when $\hbar c_l k \gg m_s c_l^2$ linear propagation dominates the spectrum in which case Eq.~(\ref{BCSspectrum2}) reduces to 
\begin{eqnarray}
E_k = \pm  \left( \hbar c_l    k  \pm  m_s c_l^2  \right) \,  ,    \label{BCSspectrum5}
\end{eqnarray}
corresponding to two copies of the Dirac spectrum shifted up or down by $\pm m_s c_l^2$. Conversely, in the strongly superconducting regime the pairing function, and hence Cooper pair mass, is large compared to the positron-electron mass. This condition reads $\Delta_p \gg m_s c_l^2$ or $2 \mu \gg m_s c_l^2$, in terms of the chemical potential of the condensate. In the limit where the kinetic energy is large as well, i.e., $\hbar c_l k \gg m_s c_l^2$, the four branches of the spectrum are gapped and Dirac-like:
\begin{eqnarray}
E_k = \pm \left(   \sqrt{5} \mu   +   \frac{ \sqrt{5}  \hbar c_l }{ 5   } k \right)   \, ,   \label{BCSspectrum6}
\end{eqnarray}
and 
\begin{eqnarray}
E_k = \pm \left(   \sqrt{3} \mu   +   \frac{ \sqrt{3} \hbar c_l  }{ 3 } k  \right)  \, .    \label{BCSspectrum7}
\end{eqnarray}
In the non-relativistic limit $v/c_l = \hbar k/m_s c_l \ll 1$, Eq.~(\ref{BCSspectrum}) reduces to two non-propagating modes 
\begin{eqnarray}
E_k = \pm  \sqrt{  ( 2 m_s c_l^2  +   \mu )^2 + 4 \mu^2    } \, ,   \label{BCSspectrum6}
\end{eqnarray}  
and two modes that correspond to the particle and holes states of standard BCS theory
\begin{eqnarray}
E_k = \pm  \sqrt{ \left( \frac{  \hbar^2 k^2 }{2 m_s} -  \mu \right)^2    + \Delta_p^2   }\, ,    \label{BCSspectrum6}
\end{eqnarray}  
where we have reinserted the superconducting gap notation $\Delta_p = 2 \mu$. The various limits and regimes are displayed in Fig.~\ref{BCS_Regimes}.

\subsection{Realization in spin-orbit coupled Bose-Einstein condensates}

Our results so far can be implemented for a spin-orbit coupled BEC by considering a 2D pseudospin-1/2 Rashba system with variable pseudospin interactions~\cite{Zhai2010}. We note that the spin-orbit coupling of the Rashba form has yet to be realized but many proposed methods exist (see for example Refs.~\cite{Zhu2006,Liu2007,Stanescu2007,Anderson2013,Xu2013}). The defining Hamiltonian $\hat{ H} = \hat{ H}_0 + \hat{ H}_\mathrm{int}$ reads
\begin{eqnarray}
\hat{ H}_0 &=& \int \! d^2{\bf r} \,  \Psi^\dagger \frac{1}{2m} \left( {\bf p}^2  + 2 \kappa {\bf p} \cdot {\bf \sigma }   + \hbar \delta \sigma_z \right)  \Psi\, ,  \label{Hzero} \\
\hat{ H}_\mathrm{int}  &=&  \int \! d^2{\bf r}  \left( g_1 \hat{n}_1^2  + g_2  \, \hat{n}_2^2 + 2 g_{12} \,   \hat{n}_1 \, \hat{n}_2 \right) \label{Hint} \, , 
\end{eqnarray}
where $\Psi = \left( \Psi_1 , \, \Psi_2 \right)^T$, ${\bf \sigma} = \left\{ \sigma_x , \, \sigma_y     \right\}$, $\hat{n}_{1 (2)}= | \Psi_{1  (2)}|^2$, $\delta$ denotes the laser detuning from Raman resonance, and $g_1$, $g_2$, $g_{12}$, are the couplings between pseudospin components. Note that the strength of spin-orbit coupling $\kappa$ depends on the relative incident angle of the Raman beams. In the present context we use the standard pseudospin notation which maps to the honeycomb lattice notation by $\Psi_{1 (2)} \to \Psi_{A (B)}$. In practice, there are several ways to eliminate the quadratic dispersion in Eq.~(\ref{Hzero}). The most straightforward approach would be to consider symmetric wavepackets with $\langle p \rangle = 0$ and momentum width $\Delta p \ll 2 \hbar \kappa, \sqrt{2 \hbar \Delta m}$, in which case one may safely neglect the ${\bf p}^2$ term in $\hat{H}_0$~\cite{Merkl2010}. A second approach is to implement a setup similar to that in Ref.~\cite{Konotop2014}. In this method, atoms are pumped from the two ground states via a complex external potential into the ${\bf p} =0$ state. The key point here is to maintain a stable population inversion given that the actual ground state is centered on a finite value of ${\bf p}$. Both approaches effectively convert $\hat{H}$ to a $\mathcal{C} \mathcal{P} \mathcal{T}$-symmetric Hamiltonian.

The underlying map that connects the spin-orbit coupled Hamiltonian to the linearized gapless RLSE requires setting the detuning to $\delta =0$ in Eqs.~(\ref{Hzero})-(\ref{Hint}), the ratio of spin-orbit coupling strength to atomic mass equal to the effective speed of light in the lattice $\kappa/m = c_l$, and the couplings $g_1 = g_2 = U  > 0$, $g_{12}= 0$. The mathematical steps of Sec.~\ref{RLSE} may then be implemented in the same way to arrive at Eq.~(\ref{finalcompact}). In contrast, the reductions in Sec.~\ref{ReducedTheories} then require a finite detuning set to $\delta = t_0$. Similarly, the Dirac-Bogoliubov-de-Gennes equations, Eq.~(\ref{DBdGE}), are derived from the spin-orbit coupled Hamiltonian by retaining a finite detuning and setting the values for the couplings $g_1$, $g_2$, and $g_{12}$ that correctly reproduce the different Lorentz invariant interactions. We illustrate this for the scalar-scalar and vector-vector interactions, i.e., Gross-Neveu~\cite{Gross1974} and Thirring~\cite{Thirring1958} models, respectively. The scalar-scalar interaction reads $\hat{ H}_\mathrm{int} = \int \! d^2{\bf r} \, U (\bar{\Psi} \Psi)^2 = \int \! d^2{\bf r} \, U \left(  |\Psi_1|^4 + |\Psi_2|^4 - 2  |\Psi_1|^2 |\Psi_2|^2 \right)$, which is obtained in Eq.~(\ref{Hint}) for $g_1 = g_2 \equiv U$, $g_{12} \equiv  - 2 U$, where $U > 0$. The vector-vector case reads $\hat{ H}_\mathrm{int} = \int \! d^2{\bf r} \, U (\bar{\Psi} \gamma^\mu \Psi)^2 = \int \! d^2{\bf r} \, U \left(  |\Psi_1|^4 + |\Psi_2|^4 + 6 |\Psi_1|^2 |\Psi_2|^2 \right)$, which requires purely repulsive interactions $g_1 = g_2 \equiv U$ and $g_{12} \equiv   6  U$. Once the particular form of the nonlinearity is constructed, either by adjusting on-site and nearest neighbor interactions for the lattice or by tuning the pseudospin interactions for the case of spin-orbit coupling, one may invoke the full results of Sec.~\ref{RLSE} through Secs.~\ref{ReducedTheories} and \ref{RelativisticBCS}. In summary, our analysis in this paper describes in detail the structure of low-energy fluctuations near a Dirac point of a honeycomb lattice or the zero-momentum point of a spin-orbit coupled BEC.

\section{Conclusion} 
\label{Conclusion}

In this article we have delineated the various constraints required for stabilizing a BEC at Dirac points of a honeycomb optical lattice. Energetically, we find that the Bose gas must be weakly interacting with excitations in the transverse direction suppressed relative to longitudinal ones. The latter condition can be implemented by using a relatively small vertical trap size. Additionally, Bloch states for the Bose gas must remain near enough to the Dirac point crystal momentum so that second-order band distortions are negligible. This condition is equivalent to the requirement that quasiparticle momenta remain much less than the Dirac point crystal momentum. Length constraints include a large quasi-2D effective healing length relative to the lattice spacing so that a continuum theory is physically sensible. Atomic and lattice parameters are related primarily by imposing the usual Landau criterion for dynamical stability, which relates the effective speed of light (lattice parameters) to the quasi-2D renormalized speed of sound (atomic parameters).

We performed a detailed analysis of lifetimes for nonlinear Dirac vortices, elucidating the low-energy landscape for each solution type. Vortex lifetimes were computed based on dynamical instabilities induced by quantum fluctuations: complex eigenvalues appear in the linear spectrum for all vortex types. These include a complex topological vortex, topological vortices with generic winding, ring-vortex, ring-vortex/soliton, vortex/soliton, Mermin-Ho, Anderson-Toulouse, and half-quantum vortices. The longest lived vortices are the ring-vortex, ring-vortex/soliton, vortex/soliton, and Anderson-Toulouse vortex with lifetimes $0.5295$ s, $4.043$ s, $3.841$ s, and $4.041$ s, respectively.

A significant part of our work was devoted to the derivation and analysis of the relativistic linear stability equations (RLSE). We demonstrated that the RLSE reduce to several well known equations. The presence of a mass gap through an offset in the sublattice potential depths allows for an interpolation between the RLSE and the BdGE. By tuning the ratio of the gap to the chemical potential between small and large values, the governing equations for quasiparticles vary continuously between RLSE and Bogoliubov-de Gennes equations (BdGE) passing through a Klein-Gordon type structure associated with fluctuations of the nonlinear Klein-Gordon equation. In the particle regime where momenta are large compared to the interaction strength, the three types of stability equations reduce to the standard Dirac and Schr\"odinger equations with the Klein-Gordon equation interpolating between these. In the single-mode approximation, where the pseudospin valley spatial functions are equal, the RLSE reduce to the Andreev equations for electrons in inhomogeneous superconductors. For zero-energy modes residing at the core of a defect such as a vortex, the RLSE reduce to the Majorana equation with the Majorana mass determined by the local density of the condensate at the ``notch'' in the case of the vortex/soliton, and equal to the chemical potential in the general case of higher winding vortices ($\ell > 1$).

By including nearest-neighbor interactions and a mass gap we have shown that the RLSE transform to the Dirac-Bogoliubov-de-Gennes equations, which describe Cooper pairing of relativistic fermions. The additional Nambu space elevates the two-spinors in two spatial dimensions to a four component object consistent with our RLSE. The non-relativistic limit is defined for quasiparticle momenta much smaller than the momentum scale set by the mass gap, in which case we recover standard BCS theory. In the analog picture the BCS pairing function is mapped to the total local condensate density, that is, the sum of squared moduli of the sublattice amplitudes. Superconductivity is strong or weak depending on the magnitude of the pairing function relative to the mass gap energy. We have shown that when the pairing function transforms as a scalar under the Lorentz group the absence of internal structure for the scalar term leaves an extra degree of freedom in the form of a vector potential. The difference in sublattice densities acts as an additional polarized vector potential acting on the pseudospin-Nambu spinor. 

Interesting research directions that extend the work presented in this article could include elevating the boson-honeycomb lattice problem to a relativistic field theory. The lowest-band approximation would still be viable provided the theory is regularized by imposing an upper momentum cutoff at the lattice scale. The various classes of Lorentz quartic interactions may be constructed by including nearest-neighbor interactions in the lattice, as we have outlined in Sec.~\ref{Reductions} of this article. It has been demonstrated that quartic interactions are fundamentally constrained by the conformal structure of all the terms of a particular relativistic Lagrangian~\cite{Alhaidari2014}. Thus, by tuning the sign and strength of nearest-neighbor interactions it may be possible to observe quantum phase transitions in the superfluid phase between different conformal theories associated with various relativistic field theories.

\ack{This material is based in part upon work supported by the National Science Foundation under grant number PHY-1306638, AFOSR grant FA9550-14-1-0287, the Alexander von Humboldt foundation, and the
Heidelberg Center for Quantum Dynamics. We acknowledge useful discussions with Ken O'Hara and Chris Weaver. }

\section*{References}

\bibliographystyle{unsrt}

\bibliography{NLDE_Vortex_Refs}

\end{document}